\documentclass{aa}

\usepackage{graphicx,xspace}

\usepackage{hyperref}
\usepackage{dirtytalk}
\usepackage[dvipsnames]{xcolor}
\hypersetup{colorlinks=true,linkcolor=blue,citecolor=blue,filecolor=blue,urlcolor=blue}

\usepackage[varg]{txfonts}
\usepackage{threeparttable}

\newcommand{\tess}{{\it TESS}}
\newcommand{\asassn}{{\it ASAS-SN}}

\begin{document} 

   \title{The optical photometric and spectroscopic periodicities of the cataclysmic variable SRGt 062340.2-265751}

   \subtitle{}

   \author{J. Brink
          \thanks{jbrink@aip.de}\inst{1,2,3,4}
          \and
                D.A.H. Buckley\inst{3,4,5}\fnmsep
          \and
                 M. Veresvarska\inst{6}\fnmsep
           \and
                A.D. Schwope\inst{1}\fnmsep
            \and 
                P.J. Groot\inst{3,4,7}\fnmsep
            \and 
               J.R. Thorstensen\inst{8}
            \and
                V. A. Cúneo\inst{1}
            \and 
                S.B. Potter\inst{3,9}
            \and
                N. Titus\inst{3}
            \and
                D. Egbo\inst{3,4}
            \and
                R. Lees\inst{3,4}
            \and
                O. Mogawana\inst{3,4}
            \and
                A. van Dyk\inst{3,4}
          }

   \institute{Leibniz-Institut für Astrophysik Potsdam (AIP), An der Sternwarte 16, 14482 Potsdam, Germany
         \and
             Institute for Physics and Astronomy, University of Potsdam, Karl-Liebknecht-Str. 24/25, 14476 Potsdam, Germany
        \and
            South African Astronomical Observatory, PO Box 9, Observatory Road, Observatory 7935, Cape Town, South Africa
        \and
            Department of Astronomy, University of Cape Town, Private Bag X3, Rondebosch 7701, South Africa
        \and
            Department of Physics, University of the Free State, PO Box 339, Bloemfontein 9300, South Africa
        \and 
            Department of Physics, Centre for Extragalactic Astronomy, Durham University, South Road, Durham DH1 3LE, UK
        \and
            Department of Astrophysics/IMAPP, Radboud University, PO Box 9010,NL-6500GL Nijmegen,the Netherlands
        \and
            Department of Physics and Astronomy, Dartmouth College, Hanover NH 03755, USA
        \and
            Department of Physics, University of Johannesburg, PO Box 524, Auckland Park 2006, South Africa
        }

   \date{Received 27 August 2025; accepted 13 January 2026}

  \abstract
   {We report on optical spectroscopic and photometric follow-up observations of the eROSITA discovered transient SRGt 062340.2-265751 and show that it displays the characteristics of a nova-like cataclysmic variable (CV), with possible indications of being a magnetic system.}
   {We try to put better constraints on the classification of SRGt 062340.2-265751 using optical time-resolved spectroscopic and photometric observations to find any periodicities in the system. From these periodicities we can classify the CV sub-type that it belongs to.}
   {Time-resolved photometric and spectroscopic observations were obtained at the South African Astronomical Observatory (SAAO), using the 1.9m and the two 1.0m telescopes. High-resolution spectra were also taken using the Southern African Large Telescope (SALT), to study the morphology of the emission lines seen in the system. Archival photometric data was also analysed in this study, including ASAS-SN, CRTS, and \textit{TESS} observations.}
   {Spectroscopic observations revealed a very low amplitude, K $\sim$ 14 km s$^{-1}$, in the radial velocity of the H$\beta$ and H$\gamma$ emission lines, suggesting that the system is likely observed at a low inclination angle. High-speed photometric observations revealed highly stochastic variability, characteristic of many magnetic cataclysmic variable systems. A probable 3.645  $\pm$ 0.006 hour orbital period was found by applying Lomb-Scargle period analysis to the H$\beta$ and H$\gamma$ emission line radial velocities. A 24.905 $\pm$ 0.065 min period was found from photometric observations, which we associate with the white dwarf spin. However, it was also found that the photometry revealed multiple periodicities from night to night. \textit{TESS} observations in three sectors did not reveal any of the periodicities found from ground-based observations, but did show a prominent period in only one sector, which might be attributed to a positive superhump period. These multiple periodicities as well as the \ion{He}{ii} $\lambda$4686 and Bowen blend emission lines seen in the spectra indicate that SRGt 062340.2-265751 is likely a nova-like CV, and might belong to the VY Scl sub-type.}
   {}


   \maketitle

\section{Introduction}

Cataclysmic variables (CVs) are binary systems, with orbital periods typically of a few hours, consisting of a white dwarf (WD) primary together with a Roche lobe filling secondary star (also called the donor or companion) that is generally of a low-mass main-sequence spectral class \citep[for a detailed description of CVs see][]{warner2003cataclysmic}. As the companion overflows its Roche lobe, material flows towards the WD through the inner Lagrangian point, \textit{L}1. The magnetic field of the WD, however, strongly dictates how the material flows as it leaves \textit{L}1. In systems with a weakly magnetized WD (B $\lesssim$10$^5$ G), an accretion disc forms that may extend all the way to the surface of the WD. In contrast, in AM Her systems, or polars, the WD is strongly magnetized (B > 10$^7$ G), the WD spin and binary periods are synchronized, and no accretion disc forms. Material is directly channelled by the magnetic field lines of the WD at \textit{L}1 \citep{schwope2025AA...698A.106S}. This causes the material to leave the orbital plane and flow directly to the WD's magnetic poles. Due to the strong magnetic field, polars have synchronized (or very nearly) WD spin and binary orbital periods. 

In CVs that contain a WD with a weaker magnetic field, known as intermediate polars (IPs) or DQ Her systems, an accretion disc may form that gets truncated at the magnetospheric radius of the WD. Material from the disc flows towards the WD in accretion curtains that rise above and/or below the orbital plane. This, together with the asynchronous WD spin and binary orbital periods, creates a rich phenomenology in periodicities in IPs. 

Disc accreting CVs are further subdivided into the low-accretion-rate dwarf novae (DNe) and the high-mass-transfer-rate nova-likes (NLs). The CVs can be further subdivided based on their spectroscopic and/or photometric properties. One of these subclasses is the VY Sculptoris (VY Scl) system, which has a distinct characteristic of going into low states at random without showing any outburst \citep[][]{hellier2001cataclysmic}. These low states are easily identified in optical light curves. It was also suggested by \citet[][]{VYScl_msgnetic} that VY Scl systems should host a magnetic WD, with similar magnetic field strengths to that found in IPs, to prevent DN outbursts during low states.

SRGt 062340.2-265751, hereafter SRGt 062340, was first reported by Denis Denisenko as a possible NL and VY Scl type variable in April 2017, and given the designation DDE 79\footnote{https://vsx.aavso.org/index.php?view=detail.top\&oid=477558}. It was later detected as a transient X-ray source \citep[][]{schwope2022identification} by both instruments, the Mikhail Pavlinsky Astronomical Roentgen Telescope - X-ray Concentrator \citep[ART-XC,][]{pavlinsky2021art}, and the extended ROentgen Survey with an Imaging Telescope Array \citep[eROSITA,][]{predehl2021erosita}, during the second all-sky survey of the Spektrum-Roentgen-Gamma \citep[SRG,][]{sunyaevSRG} mission. It was also previously detected and catalogued as a transient by The Zwicky Transient Facility (ZTF19aaabzuh), Swift/XRT (2SXPS J062339.9-265751) and ROSAT (1RXS J062339.8-265744), where it was cross-matched\footnote{https://arxiv.org/abs/astro-ph/0004053v2} to a USNO A2 optical counterpart of $B$=12.7 mag with a fractional probability of 0.952. Gaia DR3 (Gaia DR3 2899766827964264192) measured a parallax of $\pi$ = 1.9749 $\pm$ 0.0192 mas, giving a geometric distance of 495.5$\pm$4.0 pc \citep[following][]{2021AJ....161..147B}, and mean $G$ = 12.436 $\pm$ 0.005 mag, giving it an absolute G-band magnitude of $M_{G}$ = 3.96 mag, absorption not considered. Initial spectroscopic follow-up observations were made using the Wide Field Spectrograph \citep[WiFeS,][]{dopita2010wide} on the Australian National University's 2.3 m telescope, covering $\lambda$3500 - 5500 \AA, and the High-Resolution Spectrograph \citep[HRS,][]{barnes2008optical, crause2014performance} on the 10m-class Southern African Large Telescope \citep[SALT,][]{buckley2006completion}, covering $\lambda$3800 - 8900\AA\ \citep[][]{schwope2022identification}.

These spectroscopic observations revealed a continuum that is very blue. The HRS spectrum clearly shows the \ion{He}{ii} $\lambda$4686 emission line as well as the \ion{C}{iii}/\ion{N}{iii} Bowen blend (\ion{C}{iii} components at 4647.4\AA\ and 4650.1\AA, as well as \ion{N}{iii} components at 4634.13\AA\ and 4640.64\AA), a feature that is encountered in the spectra of magnetic CVs, i.e. polars and IPs, \citep[][]{schachter1991bowen, harlaftis1999emission}, and also in some nova-like CVs and low-mass X-ray binaries (LMXBs), which indicates strong UV/EUV/X-ray emission. The WiFeS observations revealed broad Balmer absorption lines with centres filled with prominent emission lines, as well as neutral and ionized helium emission lines. The full widths at half maximum (FWHMs) of the Balmer emission lines were observed to be 300-380 km s$^{-1}$, while those of the absorption lines corresponding to 3000 km s$^{-1}$. The FWHM of the \ion{He}{ii} $\lambda$4686 line was $\sim$380 km s$^{-1}$, although the line was fairly weak. A 65 km s$^{-1}$ difference in velocity was measured between the WiFeS and HRS observations, which illustrated that a follow-up orbital radial velocity study of the object was viable.

\citet{schwope2022identification} also reported on two observations of SRGt 062340 obtained with the Transiting Exoplanet Survey Satellite \cite[\textit{TESS},][]{ricker2014transiting}. The first, Sector 6, had a 30-minute cadence and was observed between 2018 December 15 and 2019 January 06. Lomb-Scargle period analysis \citep[][]{lomb1976least, 1982ApJ...263..835S} revealed strong evidence of periodicity at 3.941 $\pm$ 0.010 hours, which was attributed to be the orbital period of the system by \citet{schwope2022identification}. It was later again observed in Sector 33 with a 2-minute cadence between 2020 December 18 and 2021 January 13. A periodogram of this observation revealed maximum power at 24.37 minutes, possibly the WD rotation period, with no indication of the 3.941 hour periodicity found in the Sector 6 observation. Additional ground-based photometric observations conducted using the 1.0m telescope at the South African Astronomical Observatory (SAAO), however, did not find any of the periods detected in the \textit{TESS} observations, and instead revealed strong evidence of a 34.7-minute period, possibly indicative of an IP classification.

One of the fundamental properties of a CV is the periodicities, both orbital and WD-spin, found in such a system. Therefore, the optical time-resolved spectroscopic and photometric observations presented in this paper were obtained to put better constrains on the nature of this system, with determining these periodicities as the primary question to be answered.  The time-resolved spectroscopic observations are especially revealing, as this is the first time that such observations are published for this system. To further determine the nature of SRGt 062340, a parallel paper by \citet[][]{cuneo2026A&A...705A..71C} reports on XMM-Newton and eROSITA X-ray observations. In this paper, the results from \citet[][]{cuneo2026A&A...705A..71C} will also be used to establish the most likely periodicities found in this optical study. 
This paper is structured as follows, Section \ref{Sec:Obseravtions} gives an overview of the spectroscopic and photometric observations that were obtained, as well as the archival photometric observations used, while Section \ref{Sec:Analysis} gives a description of how the data was reduced and analysed. In Section \ref{Sec:Discussion} the results are discussed, and in Section \ref{Sec:Conclusion} the conclusions of these observations are made. The appendix consists of three sections. The first, appendix \ref{App:Obs_logs}, are the observing logs of all the SAAO spectroscopic and photometric observations analysed. In appendix \ref{App:Spectroscopy} and appendix \ref{App:Photometry} has additional plots and tables for the spectroscopic and photometric results, respectively.  

\section{Observations} \label{Sec:Obseravtions}

\subsection{SAAO 1.9m SpUpNIC spectroscopy}

Long-slit spectroscopic observations of SRGt 062340 using the SpUpNIC spectrograph \citep[][]{crause2016spupnic} on the SAAO 1.9m telescope were conducted on several epochs between 2021 February and 2021 March. All observations were made using grating 4; however, two different grating angles, $5.50^\circ$ and $4.20^\circ$, were used. The resolution of the spectra were determined by measuring the FWHM of several wavelength calibrated arc lamp spectra and is $\sim$2.3\AA\ for both configurations. A summary of these observations is given in Table \ref{tab:2SXPS_obs_log} in the appendix.

Standard \texttt{IRAF} \citep{tody1986iraf} procedures were followed in the reduction of these observations. On some nights bias and flat field images were not taken, and the calibration images from the nearest nights where those were taken was then used. Due to the shifting orientation of the spectrograph through the course of an observation, that might cause flexure in the instrumentation, calibration arc images were taken throughout the observations. The science images that were taken between two arcs were wavelength calibrated by an arc image that is the median combination of the arcs taken before and after the given science images. Spectrophotometric standard star observations were also carried out on several nights which were used to flux calibrate the observations.

\subsection{SALT HRS}

SALT HRS observations were conducted on several epochs from 2023 November - 2024 April in low-resolution mode (R $\sim$15 000) where, for each epoch, either two or four spectra were obtained. A summary of these observations is given in Table \ref{tab:2SXPS_HRS_obs_log}. HRS is a dual-beam fibre-fed \'{e}chelle spectrograph with a wavelength coverage of $\lambda$ $\sim$ 3800-8900\AA, where the `blue arm' covers $\lambda$ $\sim$ 3800-5500\AA, and the `red arm' covering $\lambda$ $\sim$ 5500-8900\AA, respectively . Initial reductions, flat field and bias corrections, were performed using the \texttt{PySALT} package \citep[][]{crawford2010pysalt}, while the spectrum was extracted using the HRS pipeline \citep{kniazev2016mn48}, which incorporates \texttt{MIDAS} \citep[][]{1983Msngr..31...26B} routines. 

\subsection{SAAO photometry}

Optical photometric observations were carried out for SRGt 062340 on 16 epochs from 2020 November 1 to 2021 March 2. Table \ref{tab:2SXPS_photometry} gives a summary of all the photometric observations. The observations were taken using the two SAAO 1.0 metre telescopes, with most of the observations being obtained with a clear filter (in white light); however, some observations were also obtained in g' and r'. Those observations taken on the nights of 2020 November 2 and 3, as well as 2020 December 3 and 4 were of poor quality and were excluded from any analysis. The adaptive elliptical aperture photometry \citep[\texttt{TEAPhot},][]{2019A&A...629A..21B} package was used to perform differential photometry, with UCAC2 20781392 (V=12.08 mag) as a comparison star. Lomb-Scargle techniques from \texttt{Astropy} \citep{2022ApJ...935..167A} as well as the general tools for astronomical time series in python \citep[\texttt{Gatspy},][]{vanderplas2015periodograms} code was used to perform period analysis of the SRGt 062340 light curves (using the \texttt{LombScargleFast} model).

\subsection{\textit{TESS} photometry}

SRGt 062340 was observed in three sectors\footnote{https://mast.stsci.edu/portal/Mashup/Clients/Mast/Portal.html}. The first at 30-minute cadence in Sector 6 from JD 2458468.2742170 (2018 December 15) to JD 2458490.0452274 (2019 January 06). It was then observed at a higher 2-minute cadence in Sector 33 from JD 2459201.7324158 (2020 December 18) to JD 2459227.5723964 (2021 January 13) and Sector 87 from JD 2460663.0380630 (2024 December 18) to JD 2460689.9405712 (2025 January 14). 

\subsection{Archival photometry}

\begin{figure}
        \centering
        \includegraphics[width = \columnwidth]{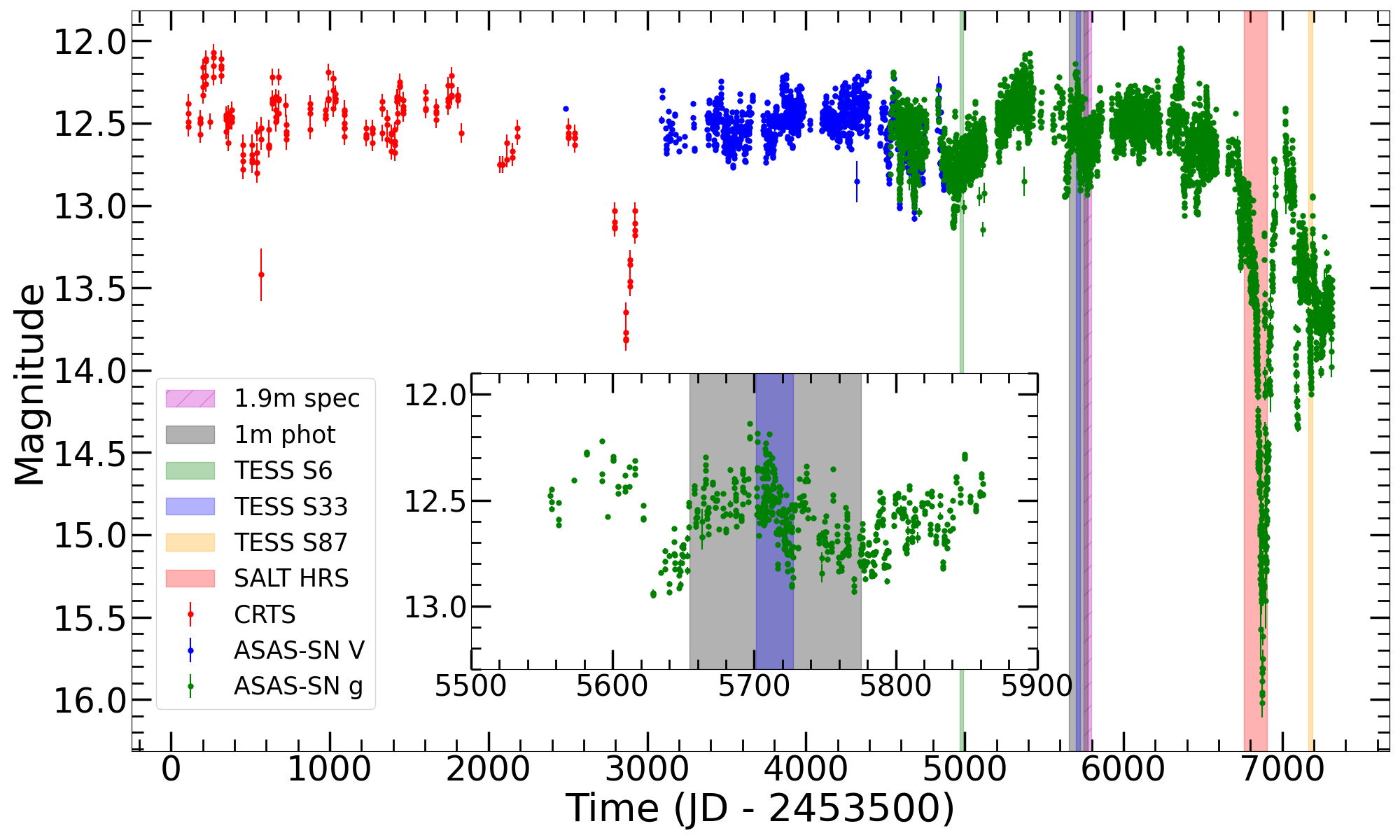}
        \caption[archival photometry]{CRTS $V$-band (Catalina ID 3027043037227) and ASAS-SN light curve of SRGt 062340 taken using g and V filters. Superimposed are the photometric and spectroscopic observations reported on in this study. Refer to the appendix and text for more detail on observation epochs. The inset plot has the same axes as the main plot, and focuses on the epochs of photometric \textit{TESS} and SAAO observations. The magnitude ranges from 12.14 - 12.93 during the SAAO photometry campaign}
        \label{fig:2SXPS_J0623_archival_photometry}
\end{figure}

Additional archival long-term photometric observations from the Catalina Real-Time Transient Survey\footnote{http://crts.caltech.edu/} \citep[CRTS,][]{drake2009first} was found that spans from JD 2453610.74097 (2005 August 28) to JD 2456422.41004 (2013 May 9). Further archival data was found from the All-Sky Automated Survey for Supernovae\footnote{https://asas-sn.osu.edu/} \citep[ASAS-SN,][]{kochanek_2017} that spans from HJD 2456596.011616 (2013 October 30) to HJD 2460815.46763 (2025 May 19). Figure \ref{fig:2SXPS_J0623_archival_photometry} shows the combined archival CRTS and ASAS-SN light curve, with the epochs of the observations reported in this paper highlighted. The inset plot focuses on the epoch of the SAAO photometric observations, and show variability in the light curve during this observing run.  

\section{Analysis} \label{Sec:Analysis}

\subsection{SAAO 1.9m long-slit spectroscopy}

\subsubsection{Spectral line measurements}

\begin{figure}
	\includegraphics[width=\columnwidth]{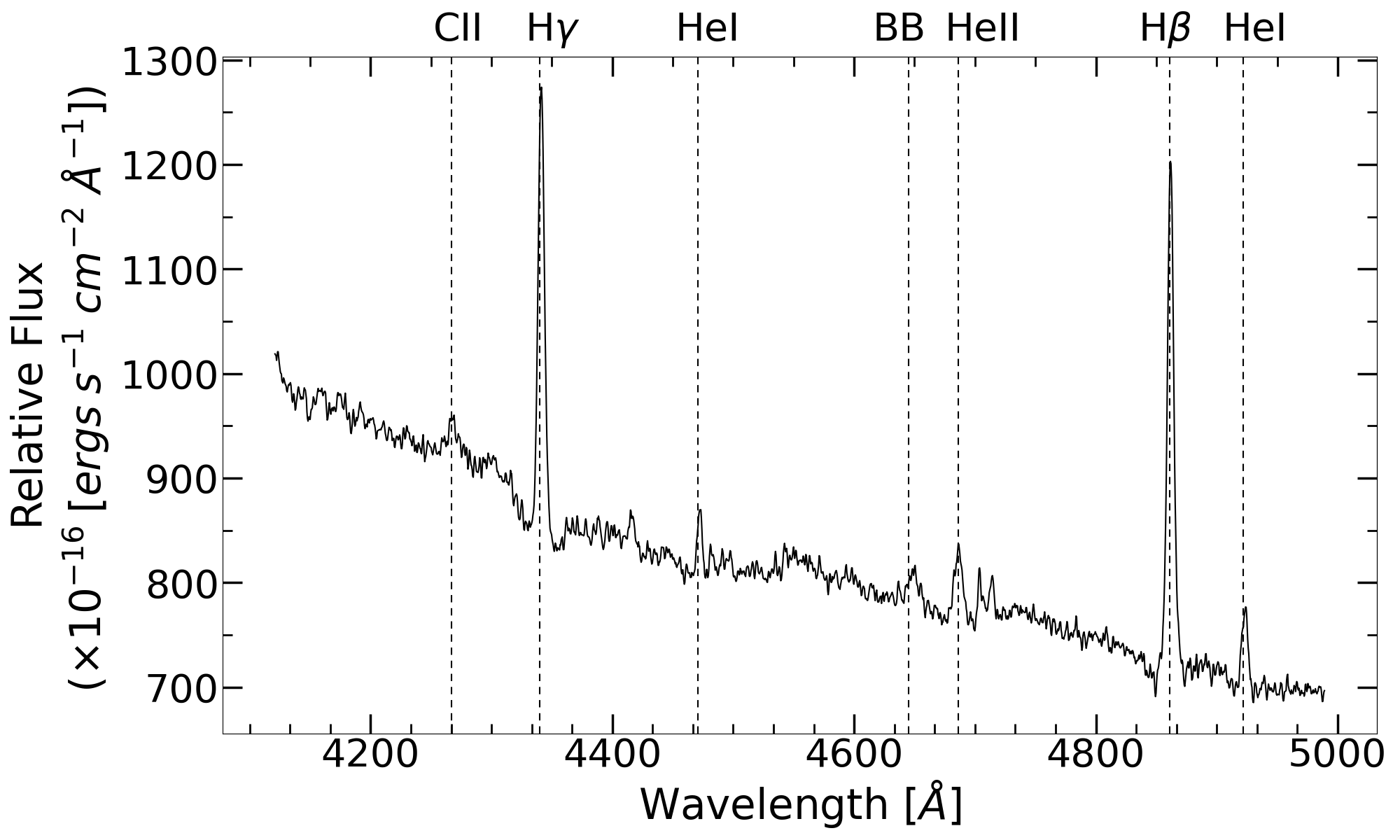}
    \caption[Spectrum]{Median combined flux calibrated spectrum ranging from 4100\AA - 5000\AA \;of the nine observations made on 2021 February 2. `BB' indicates the Bowen fluorescence blend.}
    \label{fig:SGRt_0623_spectrum}
\end{figure}

Figure \ref{fig:SGRt_0623_spectrum} shows the median combined and flux calibrated spectrum from the 2021 February 2 SpUpNIC observation, as a typical example, which shows a continuum that is rising to higher intensities towards the blue. The spectrum shows very prominent, narrow and single peaked emission lines, centred on shallow broad absorption troughs, for the Balmer H$\beta$ and H$\gamma$ lines. Other weaker emission lines are seen, such as \ion{He}{ii} $\lambda4686$ together with the Bowen blend around $\lambda$4640, as well as the \ion{He}{i} $\lambda$4471, $\lambda$4921 lines and possible \ion{C}{ii} $\lambda$4267. Figure 6 in \citep{cuneo2026A&A...705A..71C} shows the SED of the system, from which we can conclude that the absorption lines likely emanate from an optically thick accretion disc, and not from the photosphere of the WD. Donor star spectral features are usually, if they are present, seen at longer wavelengths than that covered in these observations, and hence no spectral features that could be associated with the donor star are seen in the spectrum. The spectroscopic analysis of the SpUpNIC observations will focus on the H$\beta$, H$\gamma$ and \ion{He}{ii} $\lambda$4686, being the strongest lines. Figure \ref{fig:SGRt_0623_trailed_spectra} show trailed spectra of H$\beta$ and H$\gamma$ for the February 27 observation, the longest continuous observation in our dataset. The emission lines show little change, by eye at least, in either intensity, structure or position with time, although the line centroids appear slightly shifted towards longer wavelengths compared to the respective rest wavelengths (vertical dashed lines).

Broad absorption lines are seen as the broad dark regions adjacent to the emission lines, and is most prominent for H$\gamma$. The strength of this absorption does not appear to be constant, with sporadic increases (darker bands) in absorption strength, appearing in both the Balmer lines simultaneously; however, it is slightly more pronounced for H$\beta$. The duration and frequency of these sporadic increases in absorption is also not constant. There is no indication in any of the trailed spectra of an emission-line `S-wave' that could be associated with the bright spot on an accretion disc, which might be due to insufficient resolution of the spectra coupled with the likely low inclination, as seems likely from the line morphology.

\begin{figure}
        \centering
        \includegraphics[width =\columnwidth]{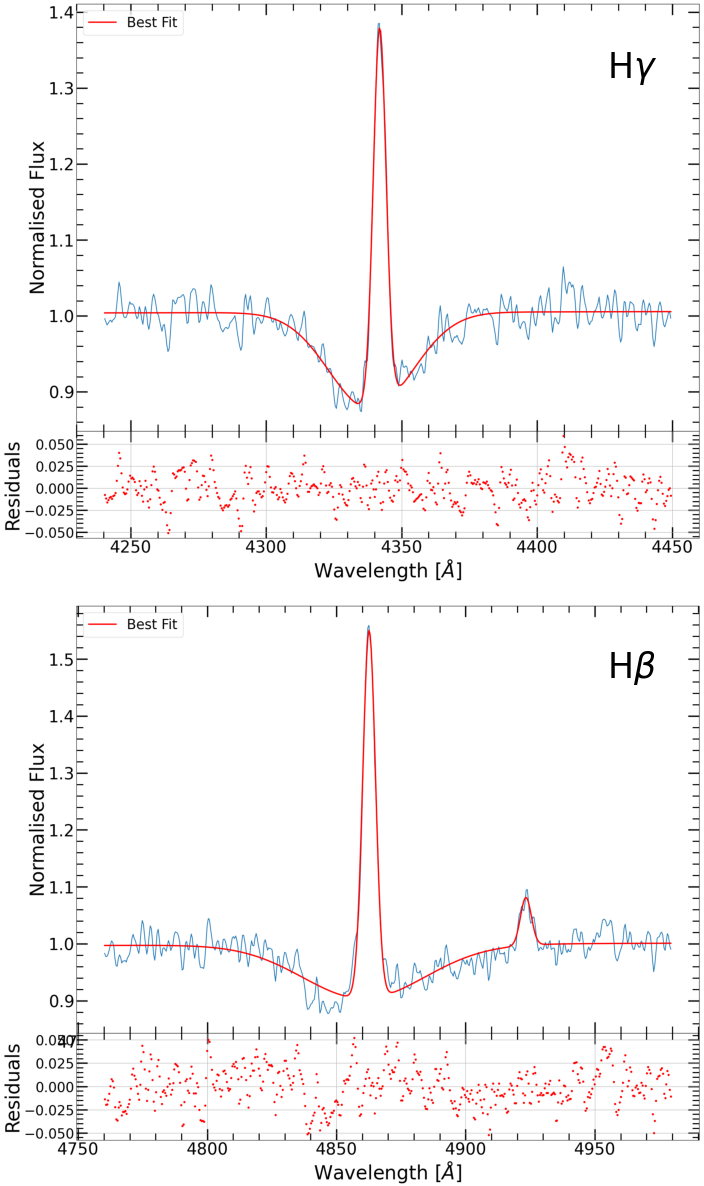}
        \caption[\texttt{lmfit} H$\beta$H and $\gamma$ fitting]{\texttt{lmfit} H$\gamma$ (top) and H$\beta$ (bottom) modelling for an observation obtained on 2021 February 26, where the bottom panels of each line shows the respective residuals of the fit.}
        \label{fig:SGRt_0623_lmfit_hgamma}
\end{figure}

To study both the emission and absorption components of the Balmer lines, the Python package \texttt{lmfit} \citep[][]{newville2016lmfit} was used to fit three components, an emission, an absorption, and one to model the continuum, to the spectra. Various different profiles were considered to model each component of the spectral line. It was found that a combination of two Gaussians, one for emission and one for absorption, together with a linear model for the continuum, had the lowest reduced $\chi^2$ value. 
This was subsequently chosen to model the spectral lines. An additional Gaussian component had to be added to the H$\beta$ model to take into account the effect of the \ion{He}{i} $\lambda$4921 line. \ion{He}{ii} $\lambda$4686 only shows an emission component and a Gaussian profile was fitted to it, as well as another Gaussian for the Bowen blend, bluewards of \ion{He}{ii}, together with a linear fit to the continuum. An illustration of the \texttt{lmfit} modelling to H$\beta$ and H$\gamma$ is shown in Figure \ref{fig:SGRt_0623_lmfit_hgamma}.

Using these model fits, parameters such as the central line positions, flux and FWHM were measured for the respective spectral lines for all the observations. Table \ref{tab:SGRt_0623_emission_line_summary} gives a summary of the averaged FWHM for the H$\beta$, \ion{He}{ii} $\lambda$4686 and H$\gamma$ emission lines per night, while Table \ref{tab:SGRt_0623_absorption_line_summary} shows the averaged H$\beta$ and H$\gamma$ absorption line measurements. The emission lines are narrow, with average widths of 316$\pm$8 km s$^{-1}$, 401$\pm$29 km s$^{-1}$, and 317$\pm$9 km s$^{-1}$ for H$\beta$, \ion{He}{ii} $\lambda$4686, and H$\gamma$, respectively. The widths of the lines do vary, with a difference between maximum and minimum width for H$\beta$ of 34 km s$^{-1}$, 20 km s$^{-1}$ for \ion{He}{ii} $\lambda$4686 and 75 km s$^{-1}$ for H$\gamma$. The Balmer absorption lines are very broad and shallow which, together with the bright emission cores, makes it difficult in determining the line parameters for the absorption components and therefore the results should be taken with caution. This is especially true for H$\beta$, as the absorption line is broader and shallower as that of H$\gamma$. 

\subsubsection{Spectroscopic orbital period}
\label{sss:spec_orb}

Lomb-Scargle period analyses were done on the H$\beta$ and H$\gamma$ emission line radial velocity measurements to find the spectroscopic period of the system. All the data apart from that obtained on 2021 February 2 was used for the analysis, as those observations were taken at a different exposure time from the rest (Table \ref{tab:2SXPS_obs_log}). The data was further divided into two sets to systematically study the period and see whether there is any change in the periodic behaviour on a scale of weeks. Data Set 1 (DS1) contains the data obtained from the 2021 February 26 - 2021 March 2 observations, while Data Set 2 (DS2) contains the data from the 2021 March 17 - 2021 March 22 observations. The respective periods found are listed in Table \ref{tab:SGRt_0623_spectroscopic_periods}. The Lomb-Scargle power spectrum for H$\beta$ DS1 has the most power at P$_{H\beta,DS1}$ = 3.650 $\pm$ 0.040 hours, while for DS2 it is at P$_{H\beta,DS2}$ = 4.341 $\pm$ 0.050 hours.
The difference in the two separate datasets for H$\beta$ corresponds to the 1 cycle/day alias as a result of the data sampling.  
The periods obtained for the two datasets of H$\gamma$ are almost identical, with P$_{H\gamma,DS1}$ = 3.646 $\pm$ 0.423 hours and P$_{H\gamma,DS2}$ = 3.659 $\pm$ 0.348 hours, respectively. 

\begin{figure}
	\includegraphics[width=\columnwidth]{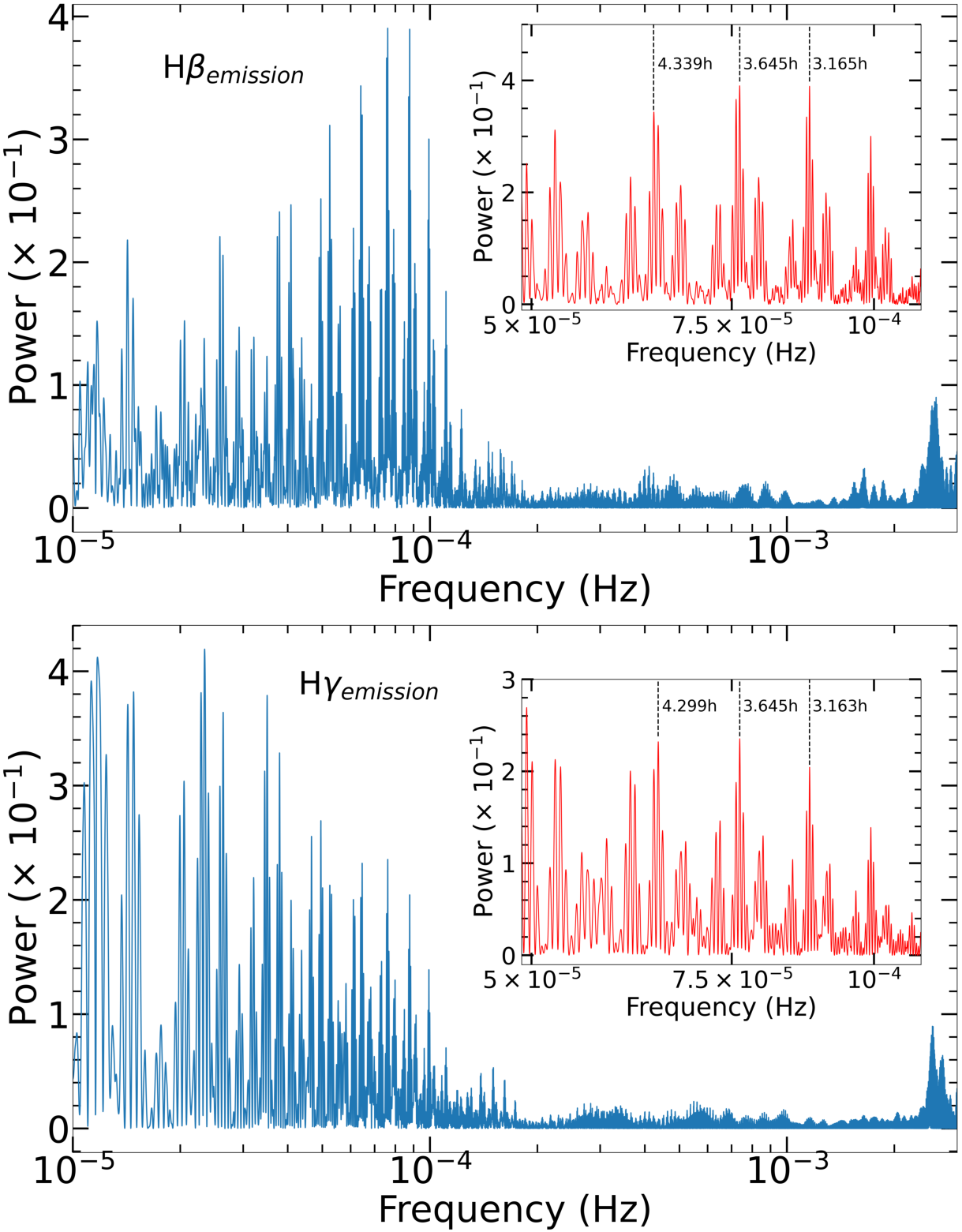}
        \caption[Emission power spectra]{Power spectra obtained using the H$\beta$ (top) and H$\gamma$ (bottom) emission line radial velocity measurements using all the datasets. The inset plot shows the frequency range of the suspected orbital period of the system. It is found that all indicated periods are one-day aliases.}
        \label{fig:SGRt_0623_RV_emission_sigma}
\end{figure}

\begin{figure}
	\includegraphics[width=\columnwidth]{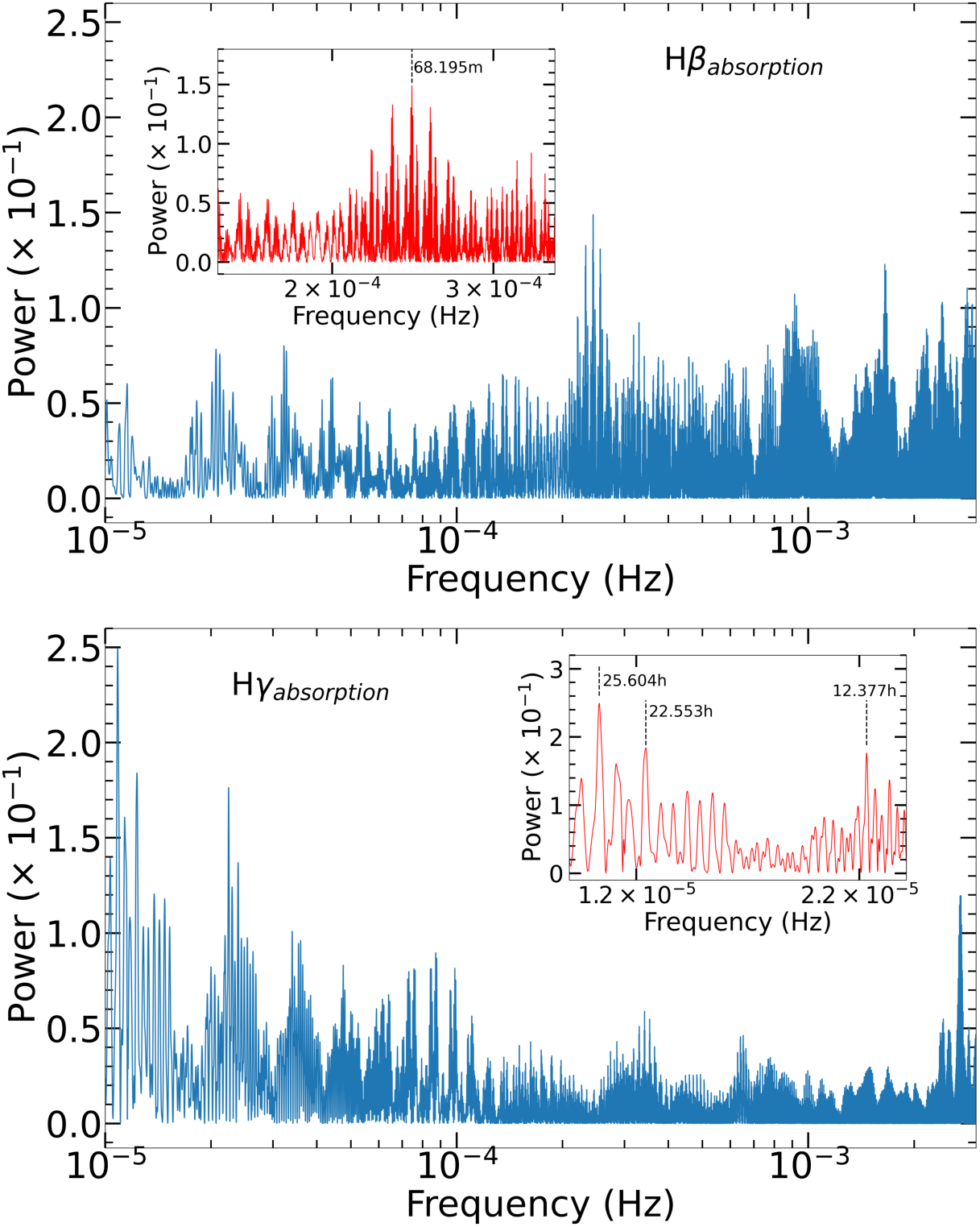}
        \caption[Absorption power spectra]{Power spectra obtained using the H$\beta$ (top) and H$\gamma$ (bottom) absorption line radial velocity measurements using all the datasets. The inset plot shows the frequency range of the suspected orbital period of the system. It is found that all indicated periods are one-day aliases.}
        \label{fig:SGRt_0623_RV_absorb_sigma}
\end{figure}

To get a better constraint on the spectroscopic orbital period, power spectra were produced by combining all (i.e. both datasets) of the H$\beta$ and H$\gamma$ emission radial velocities, respectively. Figure \ref{fig:SGRt_0623_RV_emission_sigma} show both the H$\beta$ (top panel) and H$\gamma$ (second panel) power spectra, where the inset plots show zoomed in regions around the identified periods. It is unclear why so much power is seen at low frequencies in the combined H$\gamma$ power spectrum, which has its strongest peak at a frequency that corresponds to a period of 11.86 hours. This is surprising given that the H$\gamma$ DS1 and DS2 dominant periods were both $\sim$3.65 hours. Using the DS1 and DS2 periods as a guide, as well as the periods found in the H$\beta$ power spectra, only periods below 4.5 hours were considered as probable orbital periods. Several strong peaks are seen for both H$\beta$ and H$\gamma$ in the $\sim$0.05 - 0.10 mHz frequency range, which are due to aliasing. The strongest peak in H$\beta$ is seen at 0.0762 mHz, corresponding to a period of 3.645 hours. This is also seen for H$\gamma$, from which we conclude that this is the orbital period.

\begin{table}
\centering 
\caption{Spectroscopic periods} 
\label{tab:SGRt_0623_spectroscopic_periods}
\begin{threeparttable}
\begin{tabular}{lccc}
\hline
\hline
            &   Data    & H$\beta$ Period      & H$\gamma$ Period       \\
            &           &  (h)                 &  (h)                   \\
    \hline
            &   Data Set 1   & 3.650 $\pm$ 0.040   &  3.646  $\pm$ 0.423    \\
Emission    &	Data Set 2   & 4.341 $\pm$ 0.050   &  3.659  $\pm$ 0.348    \\
            &   All          & 3.645 $\pm$ 0.006   &  3.645  $\pm$ 0.006  \\
	\hline
            &	Data Set 1   & 1.137 $\pm$ 0.041   &  3.178 $\pm$ 0.320 \\
Absorption  &	Data Set 2   & 0.173 $\pm$ 0.001   &  5.338 $\pm$ 0.748 \\
            &	All          & 0.261 $\pm$ 0.001   &  3.677 $\pm$ 0.055 \\
	\hline
	\hline    
\end{tabular}
\begin{tablenotes}
    \item Data Set 1: 2021 February 26, 27, March 2
    \item Data Set 2: 2021 March 17, 18, 19, 20, 21, 22
    \item All: All data (i.e. Data Set 1 + Data Set 2)

    \end{tablenotes}    
\end{threeparttable}
\end{table}

A similar analysis was attempted for the H$\beta$ and H$\gamma$ absorption radial velocity measurements, with the results summarized in Table \ref{tab:SGRt_0623_spectroscopic_periods} and Figure \ref{fig:SGRt_0623_RV_absorb_sigma}. 
It is immediately clear that for the H$\beta$ absorption component, there is no clear indication of the orbital period, or indeed any period above the noise threshold, apart for a 5-$\sigma$ signal at 68.195 minutes (and its corresponding aliases). This is likely because the H$\beta$ absorption lines are much shallower and broader than in the case of H$\gamma$, and therefore more difficult to fit and obtain a reliable centroid. For the stronger H$\gamma$ absorption, however, the combination of both datasets, i.e. `All', agrees within the uncertainty with the period found from the emission line radial velocities, while the Data Set 1 results is a 1-cycle per day alias of this period (Table \ref{tab:SGRt_0623_spectroscopic_periods}). 

Figure \ref{fig:SGRt_0623_rv_power_density_spectra_all_data} shows log-log power spectra of the Balmer emission line radial velocities, where the positions of the spectroscopic orbital frequency, $\Omega$ (obtained in Section \ref{sss:spec_orb}), as well as the possible spin frequency of the WD, $\omega$, are also shown (see Section \ref{subsec:saao_photometry}). Clearly there is no evidence of significant power in the radial velocities at the photometrically derived spin frequency. The power excess, or `arching effect' at $\nu < 10^{-4}$Hz, seen in the bottom panel of Figure \ref{fig:SGRt_0623_rv_power_density_spectra_all_data} is due to data gaps in the SAAO photometry.

\subsubsection{Orbital ephemeris}\label{subsec:orbital_ephemeris}

The orbital phases for each observation were determined using the spectroscopic orbital period, $P_{orb}$ = 3.645 h. This was accomplished by first arbitrarily assigning the BJD of the first observation taken on 2021 February 26 as phase 0, and then subsequently determining the phase of all the other observations from this point of reference. 
A sine function of the form $K\sin (2\pi (ft+\phi))+\gamma$, where $K$ is the amplitude, $\phi$ the phase shift, and $\gamma$ the systemic velocity of the system, was fitted to the 
H$\beta$ radial velocities to determine the phase shift that was needed to determine the BJD T$_0$, where the system shows a radial velocity exactly midway between maximum and minimum (i.e. the point of inferior conjunction of the companion star).

The fit revealed a very low amplitude of $K$ = 14.05 $\pm$ 0.67 km s$^{-1}$ and a systemic velocity of $\gamma$ = 37.34 $\pm$ 0.47 km s$^{-1}$. Applying the phase shift, determined from the sine function, the true BJD T$_0$ was derived.
The ephemeris of SRGt 062340, as determined from the H$\beta$ radial velocity measurements, is therefore
                \begin{eqnarray*}
                   T_0 = 2459273.078516(28) + 0.151875(25) \times E.
                \end{eqnarray*}

\begin{figure}
        \centering
        \includegraphics[width=\columnwidth]{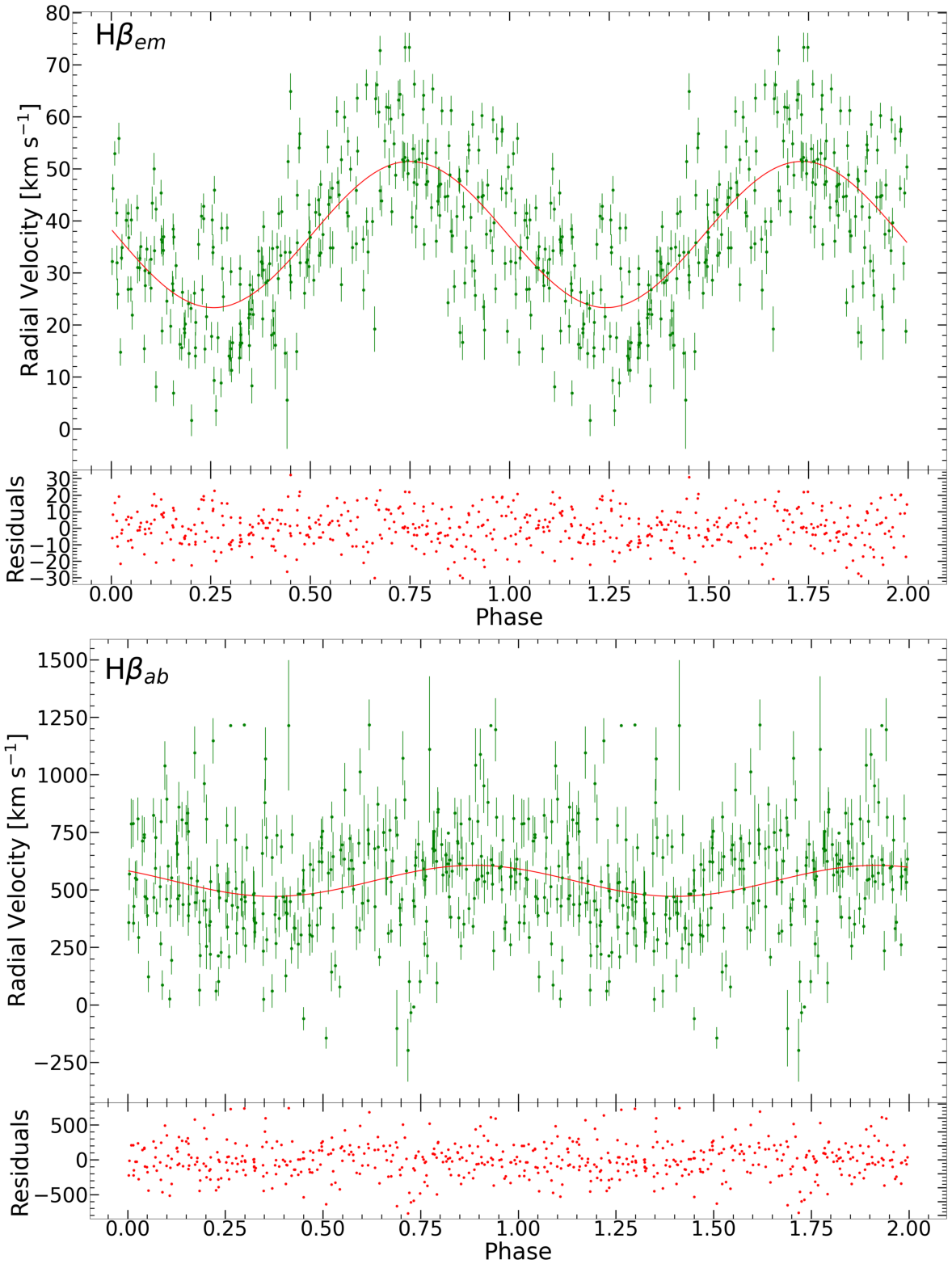}
        \caption[phase-folded H$\beta$ emission and absorption line radial velocities]{H$\beta$ emission (top) and absorption (lower) line radial velocities phase-folded on the 3.645-hour period. The fitted sine functions, with the parameters in Table  \ref{tab:SGRt_0623_phase_folded_parametrs}, are also shown, with the residuals in the lower panels of each plot.}
        \label{fig:phase_folded_h_beta_rv_lines}
\end{figure}

\begin{figure}
        \centering
        \includegraphics[width=\columnwidth]{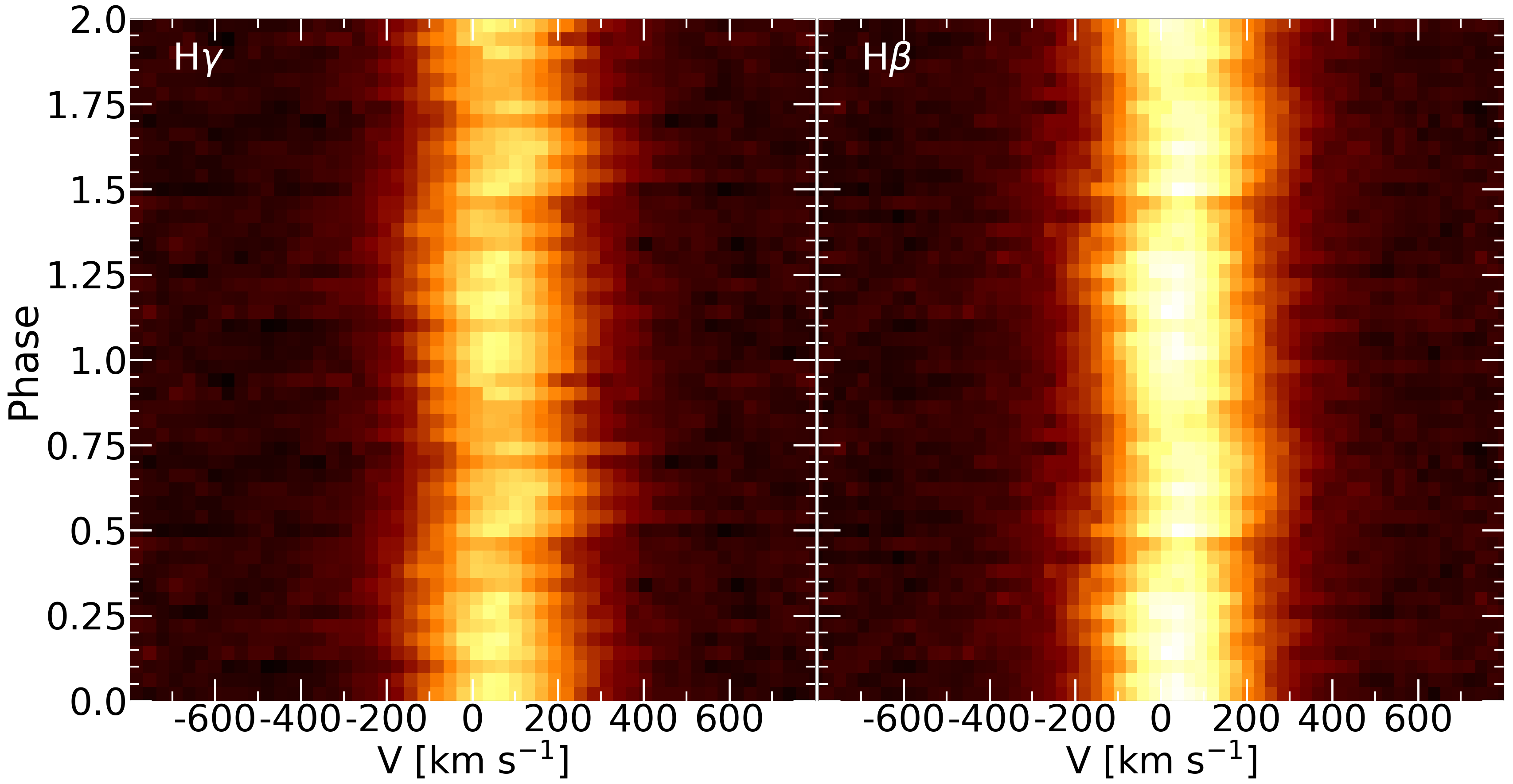}
        \caption[phase-folded trail spectra]{Trailed spectra phase-folded on the best (H$\beta$) spectroscopic orbital period (3.645h). Left is H$\gamma$ and right H$\beta$.}
        \label{fig:SGRt_0623_phase_folded_trail_spectra}
\end{figure}

Figure \ref{fig:phase_folded_h_beta_rv_lines} shows the resulting sine fit of the radial velocity measurements to the phase-folded H$\beta$ emission and absorption lines, together with the residuals of the fit. Figure \ref{fig:SGRt_0623_phase_folded_emission_absorption_rv} in the appendix show the fit to the H$\gamma$ emission and absorption lines, as well as \ion{He}{ii} $\lambda$4686 emission line. The parameters for these fits are presented in Table \ref{tab:SGRt_0623_phase_folded_parametrs}. Despite the good quality of the spectral fits, with reduced $\chi^2$ $\approxeq$ 1, there is a lot of scatter in the radial velocity measurements that are larger than the uncertainties in the individual measurements. A possible explanation for this could be that using the generic templates in the \texttt{lmfit} modelling is not sufficient for this system, and employing physically self-consistent accretion disc line profiles might, in principle, improve the stability of the radial velocity measurements. Given the available data, we however assume that the overall radial velocity behaviour is nonetheless reliably represented by our analysis. We therefore focus on the mean radial velocity behaviour, which we assume to be robust. Figure \ref{fig:SGRt_0623_phase_folded_trail_spectra} shows the phase-folded trailed spectra of H$\beta$ and H$\gamma$ in velocity space. The system has a small velocity amplitude; however, the orbital modulation is evident in the phase-folded trailed spectra. While it is not immediately clear as to why some data points appear to be outliers, one possible interpretation could be that they are due to flaring events in the system which might introduce peculiar velocities.

\begin{table}
\centering 
\caption{Phase-folded parameters} 
\label{tab:SGRt_0623_phase_folded_parametrs}
\begin{threeparttable}
\begin{tabular}{cccc}
\hline
\hline
      Line & K             & $\gamma$      & $\phi$ \\
           & (km s$^{-1}$) & (km s$^{-1}$) & \\ 
    \hline
	Emission           &                   &                &                   \\ 
    H$\beta$           & 14.05 $\pm$ 0.67  & 37.34 $\pm$ 0.47 & 0.98 $\pm$ 0.09 \\
    H$\gamma$          & 15.10 $\pm$ 1.30  & 38.46 $\pm$ 0.91 & 1.05 $\pm$ 0.17 \\
	\ion{He}{ii} $\lambda$4686 & 15.59 $\pm$ 2.71  & 47.09 $\pm$ 2.06 & 0.65 $\pm$ 0.36 \\
                       &                   &                &                   \\
    Absorption         &                   &                &                   \\ 
    H$\beta$           & 67.52 $\pm$ 14.99 & 538.90 $\pm$ 11.26  & 0.77 $\pm$ 0.44\\
    H$\gamma$          & 47.94 $\pm$ 12.97 & 27.36 $\pm$ 9.35    & 0.74 $\pm$ 0.52\\
    \hline
	\hline
\end{tabular}
\end{threeparttable}
\end{table}

\subsubsection{Phase-folded radial velocities} \label{subsec:results_spec_2sxps_phase_folded_rv}

From Table \ref{tab:SGRt_0623_phase_folded_parametrs} it is seen that all of the emission line phase values agree with each other within the uncertainties of the measurements. 
The radial velocity amplitudes for the two Balmer and \ion{He}{ii} $\lambda$4686 emission lines are all similar, with K$\sim$14.5 km s$^{-1}$, although the \ion{He}{ii} plot has a lot more scatter, being a weaker line (see Figure \ref{fig:SGRt_0623_phase_folded_emission_absorption_rv}) with consequently larger velocity uncertainties.
 
Despite the larger uncertainties associated with the absorption line measurements, compared to the emission lines, it is still instructive to compare the phase-folded absorption line radial velocities to those of the respective emission lines. 
As expected, the absorption line measurements show a lot more scatter compared to the emission lines. However, an orbital modulation in the absorption lines is still observed and it is only slightly out of phase with the emission components ($\Delta\phi \sim$ 0.2, see Table \ref{tab:SGRt_0623_phase_folded_parametrs}), though the uncertainties are large. One peculiarity 
is that of the H$\gamma$ absorption line has a $\gamma$ velocity (i.e. systemic) similar to that found for all the emission lines, whereas the H$\beta$\ line has a substantial $\gamma$ velocity of $\sim$540 km s$^{-1}$. This could be attributed to poor modelling of the H$\beta$ absorption component, which is due to the shallow and broad nature of the line with a superposed strong emission line in its core. Thus, the H$\beta$ $\gamma$ velocity should not be considered in the rest of the analysis.

\subsection{SALT HRS}\label{subsec:SALT_HRS}

\begin{figure*}
        \centering
        \includegraphics[width = 1.0\textwidth]{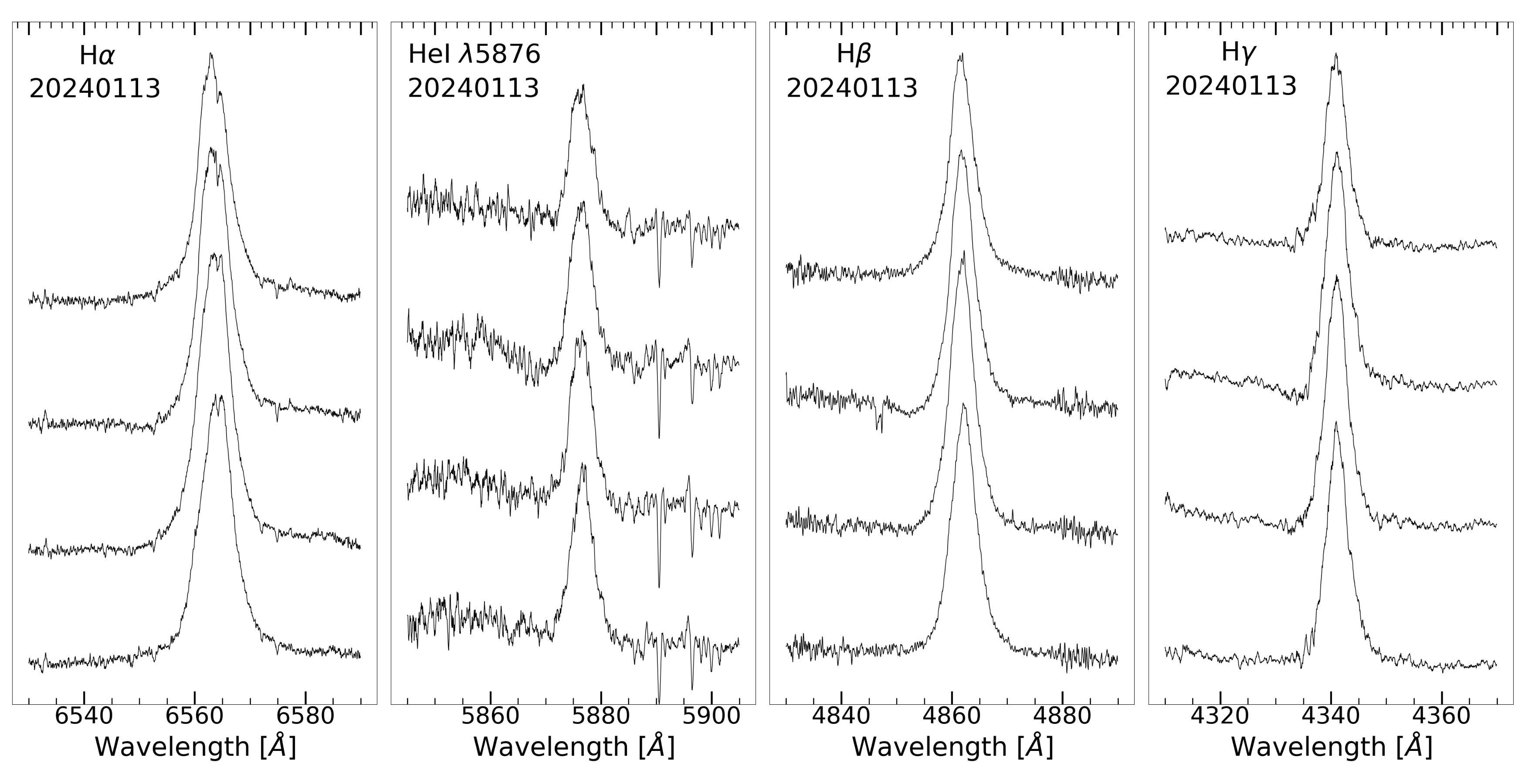}
        \caption[HRS stacked spectra]{2024 January 13 continuum subtracted normalized SALT HRS stacked spectra, focusing on the Balmer emission lines as well as \ion{He}{i} $\lambda$5876. All the spectra were smoothed by a moving 1D box kernel of size 3, 7, 4, and 11 for H$\alpha$, \ion{He}{i} $\lambda$5876, H$\beta$, and H$\gamma$, respectively. Each spectrum is a 600-second exposure, with the first spectrum being the first from the bottom and the last spectrum obtained at the top. The absorption features seen at around 5890\AA\ and 5896\AA\ in the \ion{He}{i} plot are from telluric lines. Note, the absorption to the blue in the third spectrum of the emission lines. The offset of the stacked spectra were 0.0023, 0.001, 0.0025, and 0.007 for H$\alpha$, \ion{He}{i} $\lambda$5876, H$\beta$, and H$\gamma$, respectively.}
        \label{fig:2SXPS_J0623_HRS_stacked_spectra_example}
\end{figure*}

SALT HRS observations were obtained with the aim of determining whether the Balmer emission lines were truly a simple Gaussian, as the SAAO 1.9m observations suggested, or whether the emission lines have a more complex, non-Gaussian, morphology. Figure \ref{fig:2SXPS_J0623_HRS_stacked_spectra_example} shows zoomed-in regions for H$\alpha$, \ion{He}{i} $\lambda$5876, H$\beta$, and H$\gamma$ of the four spectra that were obtained on 2024 January 13. Even though moving 1D box kernel smoothing was applied to all spectra (with sizes 3, 7, 4, and 11 for H$\alpha$, \ion{He}{i} $\lambda$5876, H$\beta$, and H$\gamma$, respectively (due to the different resolution at different wavelengths), it is evident, for H$\alpha$ at least, that the emission line does not have a strict Gaussian profile, having double peaks. These peaks seem to be changing, with both blue and red peaks being equally prominent in the first observation and the blue peak becoming more prominent in subsequent spectra. Even though the double-peak structure is not strong, it is an indication of a possible disc in the system that is seen at a low inclination angle. It is not clear whether double peaks can be seen for H$\beta$ or H$\gamma$; however, the emission line profile for all three Balmer lines does seem to be a non-uniform Gaussian, with an extended red-wing, which could indicate a high velocity component. \ion{He}{i} $\lambda$5876 also suffer from high noise levels; however, the profile appear to be Gaussian. Due to the high noise level, it is difficult to determine whether multiple components are present in the \ion{He}{i} $\lambda$5876 emission, but it does appear likely. There are indications of blueshifted absorption in the second and third spectra of Figure \ref{fig:2SXPS_J0623_HRS_stacked_spectra_example}, especially for \ion{He}{i} $\lambda$5876, which could indicate the presence of a P-Cygni profile caused by a disc wind \citep[e.g.][]{Cuneo_disc_wind}.

\begin{figure}
        \centering
        \includegraphics[width = \columnwidth]{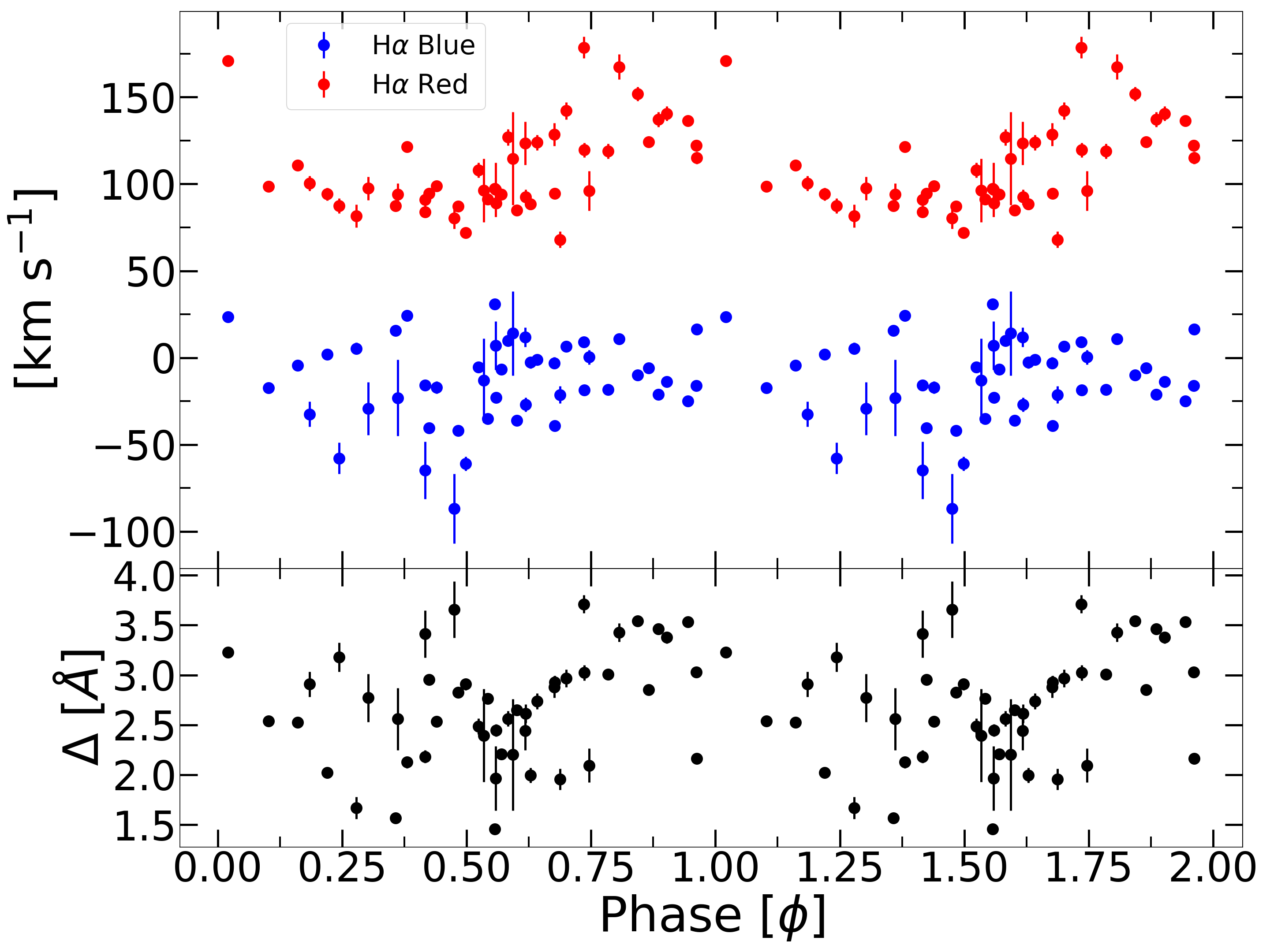}
        \caption[HRS phase RV]{Radial velocity measurements of the SALT HRS observations for both the blue and red components of H$\alpha$, phase-folded on the suspected orbital period of P$_{\text{orb}}$ = 3.645 hours (top panel). Difference in wavelength between the red and blue components (bottom panel).}
        \label{fig:2SXPS_J0623_HRS_rv}
\end{figure}

The question now arises of whether the change seen in the morphology of the H$\alpha$ emission line is phase dependent. To answer this, \texttt{lmfit} modelling to each spectrum was done using a double Lorentzian profile (which gave a better reduced $\chi^{2}$ fit to the majority of the spectra compared to a double Gaussian model), to fit a profile to the blue and red components, respectively. 
The radial velocity of both components was determined from this fit, and is captured in Table \ref{tab:appendix_2SXPS_J0623_HRS_rv}, together with the HJD and phase of the respective observations. The phase of each HRS observation was determined using the orbital ephemeris as stated in Section \ref{subsec:orbital_ephemeris}. The top panel of Figure \ref{fig:2SXPS_J0623_HRS_rv} shows the radial velocities of the blue and red components for all 48 HRS spectra, while the bottom panel of the figure shows the difference in wavelength between the two components. The radial velocities seem to roughly follow a sinusoidal pattern of very low amplitude (few tens of km s$^{-1}$), indicative of motion about the centre of mass in the system. However, the bottom panel of Figure \ref{fig:2SXPS_J0623_HRS_rv} shows a random scatter, and hence no clear phase-dependent behaviour in the morphology of the H$\alpha$ emission line.

Using a phase bin of $\phi$=0.1, spectra were phase-binned together (by median combining all the spectra that fall within a given bin using \texttt{IRAF}) to determine whether any underlaying phase-dependant morphology could be seen in the H$\alpha$ emission line. However, no phase-dependant morphology in the line was seen. As with the individual spectra, \texttt{lmfit} was used to model the binned spectra, which is summarized in Table \ref{tab:2SXPS_J0623_HRS_rv_phase_folded}.

\subsection{SAAO photometry}\label{subsec:saao_photometry}

\subsubsection{Light curves}

Strong short timescale variability (on timescales of minutes) is seen in all the light curves, with the frequency and amplitude of peaks and dips varying from epoch to epoch, with no apparent repeating pattern or frequency (confirmed in Section \ref{subsubsec:saao_phot_powerspectra} from power spectra). Figure \ref{fig:2SXPS_J0623_lightcurve_set} in the appendix show the optical light curves for each night, together with the respective Lomb-Scargle power spectrum. The light curves show the normalized differential magnitude, where the magnitude of the comparison star was subtracted from the target star, giving the differential magnitude, after which the data were normalized by subtracting the average of this differential magnitude from each data entry (with negative values being brighter than positive values). Dips generally appear to have a similar steep slope during rise and fall. Phase folding light curves on the spectroscopic orbital period (i.e. 3.645 hours) reveal that the variability does not appear to be phase dependant either, and show strong stochastic variability from orbit to orbit. Such stochastic variations, at relatively high amplitudes, are quite common in CVs and indeed many IPs (e.g. for TX Col, see Fig. 10 in \cite{1989ApJ...344..376B}).\\

\subsubsection{Power spectra}\label{subsubsec:saao_phot_powerspectra}

Table \ref{tab:2SXPS_J0623_differential_photometry_all_periods} lists all the strongest peaks found from the Lomb-Scargle period analysis of each observation epoch and shows that no single coherent period is common to all the individual observations. The uncertainties in the periods found from the Lomb-Scargle power spectra were obtained using bootstrapping techniques, similar to what was used by \cite{Zurek2009ApJ...699.1113Z}. For each light curve, 10,000 synthetic light curves were generated. A lomb-Scargle period search was again performed for each of those, finding the best period. The uncertainty in the period was then taken as the standard deviation of the best periods found of the 10,000 synthetic light curves. 
In Table \ref{tab:2SXPS_J0623_differential_photometry_combined_periods} observations from different epochs were grouped together in the hope that the longer observation baseline will put a better constraint on the photometric periods. The grouping of data was largely determined from temporal considerations, the details are given in the footnote of Table \ref{tab:2SXPS_J0623_differential_photometry_combined_periods}.  
There is again no consistent period found between the datasets and the uncertainties are large, with the exception of Data Set 6 (DS6), namely all the combined clear filter data. The larger baseline of this dataset gives the best resolution power spectra. Apart from the region of lowest frequencies near the orbital period, a strong peak was found at a much higher frequency of 0.66920 $\pm$ 0.0036 mHz (or $P$ = 24.905 $\pm$ 0.065 minutes). This is very close to two periods identified in the \textit{TESS} Sector 33 (high cadence) data power spectra, namely at 24.37 and 25.20 min, identified by \cite{schwope2022identification} as a possible spin period. This peak in the power spectrum might be interpreted as the likely spin frequency of the WD, $\omega$. 

\begin{figure}
        \centering
        \includegraphics[width = \columnwidth]{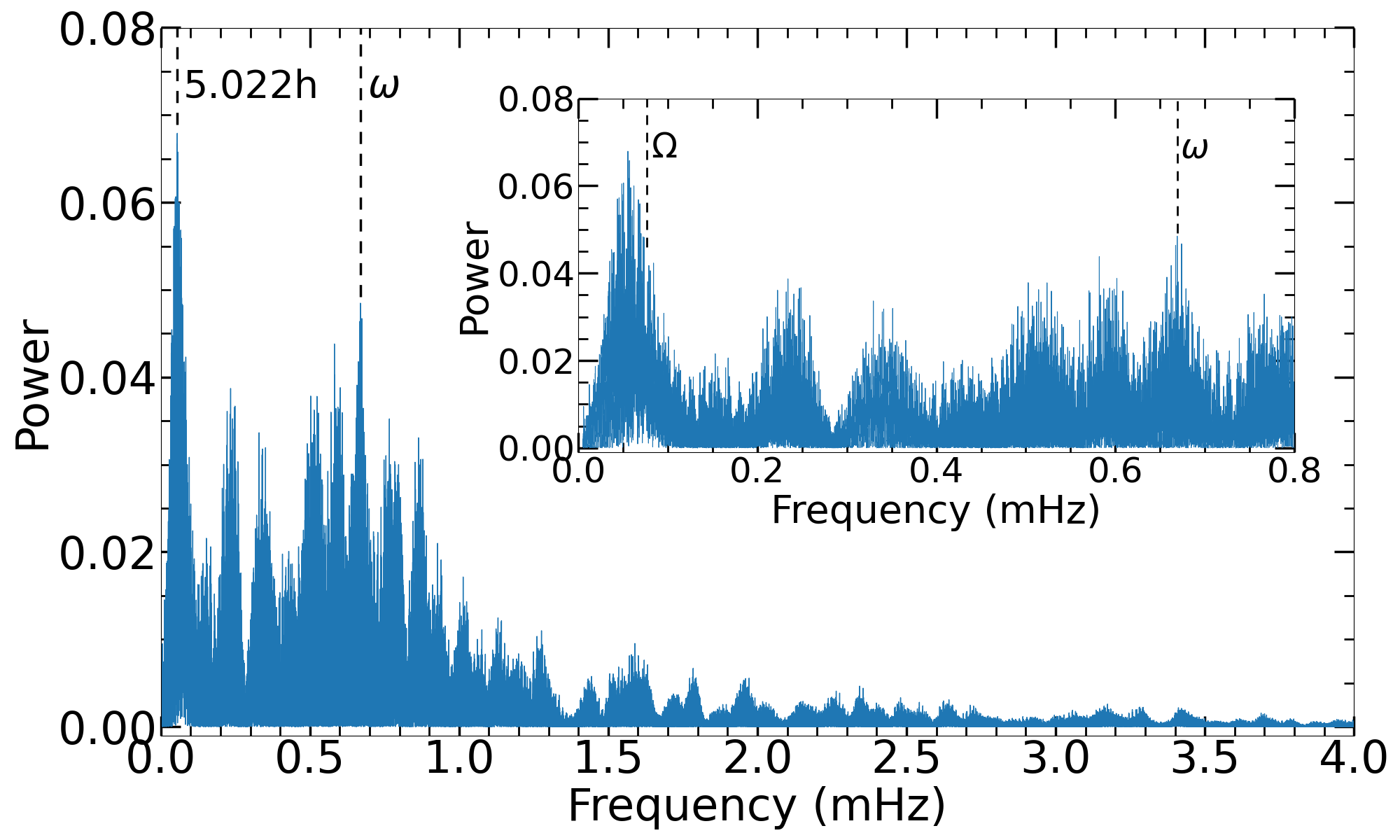}
        \caption[photometry power spectrum]{Lomb-Scargle power spectrum using all the clear filter photometric observations. The frequency of the strongest peak coincides with a period of 5.022 hours. The probable spin frequency of the WD is indicated by the dashed line labelled $\omega$. The inset plot focuses on the region 0.0 - 0.8 mHz and show the position of the spectroscopic orbital period, $\Omega$, and the probable spin frequency.}
        \label{fig:2SXPS_J0623_photometry_power_spectrum}
\end{figure}

Figure \ref{fig:2SXPS_J0623_photometry_power_spectrum} shows the resulting power spectrum for all the clear filter combined data (DS6). There is a strong indication of a peak at the rotation frequency, $\omega$, while power is seen at lower frequencies. The spectroscopic orbital frequency, $\Omega$ (associated with P$_{\text{orb}}$ = 3.645 hours), is indicated as well, but does not coincide with the strongest peak at low frequencies (which is at $\sim$5.022 hours). The inset shows a region zoomed-in on 0.4$-$0.8 mHz, showing the spin frequency. 
Figure \ref{fig:2SXPS_J0623_photometry_power_spectrum} also shows a lot of aliasing, which is due to the irregular intervals between observations within the $\sim$17 weeks of observations. 

\begin{figure}
        \centering
        \includegraphics[width = \columnwidth]{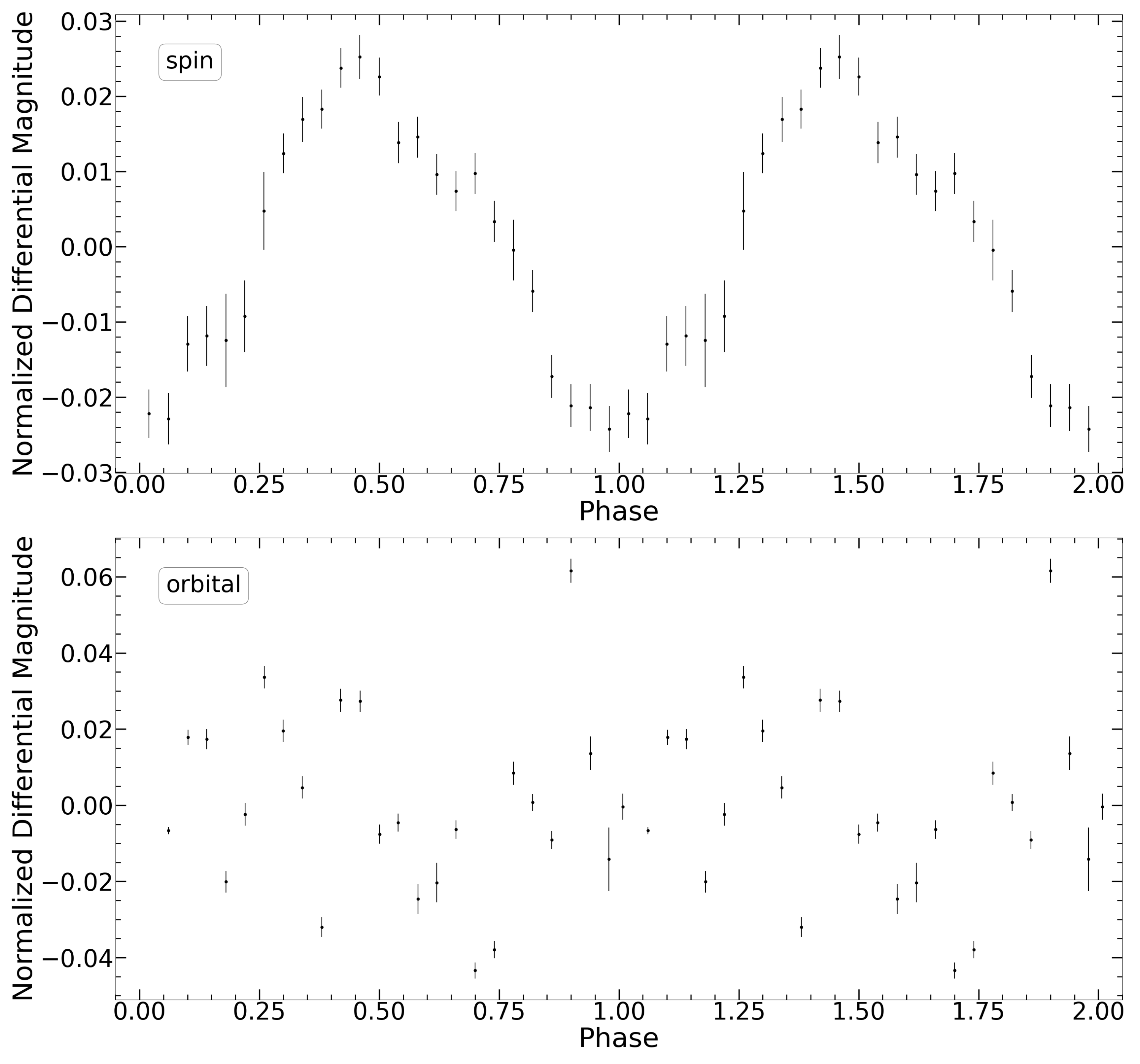}
        \caption[Binned phase-folded Photometry]{Binned phase-folded light curves of all the clear filter observations. Top panel is folded on the spin period, 24.905 minutes, while the bottom panel is folded on the 3.645-hour orbital period.}
        \label{fig:2SXPS_J0623_phase_folded_photometry_binned}
\end{figure}

Figure \ref{fig:2SXPS_J0623_phase_folded_photometry_binned} shows the clear filter observations phase-folded on the spin and orbital periods, where the observations were binned in $\phi$=0.04 bins. Modulation on the spin period is clearly seen.  

To investigate whether $\omega$ is significant and a stable period, the data was divided into three groups and subjected to period analysis. Figure \ref{fig:2SXPS_J0623_month_photometry_power_spectrum} shows the resulting power spectra of combining the observations obtained in 2020 November, 2021 January, and 2021 February. $\omega$ is also indicated in the respective plots and show that some power, and at roughly the same intensity, is seen in all three groupings at this frequency. However, the aliasing is quite complex, and it is clear that there are significant changes in the power spectra and where the maximum power is seen in all three groupings. This indicates changes on timescales of weeks, or shorter, which could imply that SRGt 062340 is subject to QPOs, similar to what has been seen in other systems, such as the IP, TX Col \citep[][]{littlefield2021quasi}. A more definitive answer to the question of whether there's a coherent spin period needs time series X-ray observation, which is reported in \citet[][]{cuneo2026A&A...705A..71C}.

\begin{table}
\centering 
\caption{Differential photometry periods when combining data sets} 
\label{tab:2SXPS_J0623_differential_photometry_combined_periods}
\begin{threeparttable}
\begin{tabular}{cccc}
\hline
\hline
Data    & Filter    & Period (hours) & Period (min)  \\
    \hline
	Data Set 1   & clear    &  0.5728 $\pm$ 0.0023 & 34.3 $\pm$ 0.1  \\
	Data Set 2   & clear	&  5.0183 $\pm$ 0.0047 & 301.2 $\pm$ 0.3  \\
	Data Set 3   & clear	&  1.3355 $\pm$ 0.0029 & 80.0 $\pm$ 0.2   \\
	Data Set 4   & clear	&  3.3274 $\pm$ 0.0185 & 202.0 $\pm$ 1.1  \\
	Data Set 5   & $g'$	    &  3.9847 $\pm$ 0.0022 & 238.9 $\pm$ 0.1  \\
	    \hline
    Data Set 6   & clear    &  0.4151 $\pm$ 0.0011 & 24.905 $\pm$ 0.065  \\
	\hline
	\hline
\end{tabular}
\begin{tablenotes}
    \item Data Set (DS) 1: 2020 November 27, 28, 29
    \item DS 2: 2021 January 4, 13, 20
    \item DS 3: 2021 February 1, 2 
    \item DS 4: 2021 February 26, 27 
    \item DS 5: 2021 February 27, March 2 
    \item DS 6: All clear observations
    \end{tablenotes}
\end{threeparttable}
\end{table}

\subsection{\textit{TESS} photometry}\label{subsec:TESS_photometry}

\citet[][]{schwope2022identification} report on the results of two \textit{TESS} observations, observed in Sector 6 and Sector 33. In this section some of those results are reproduced, and we report new \tess\ photometry from Sector 87 at 2-minute cadence. The low cadence 30-minute Sector 6 light curves were obtained from the \tess\ full-frame images, while the higher cadence 2-minute Sector 33 light curves were obtained from Pre-search Data Conditioning (PDCSAP) fluxes. Full details of the reduction of the two datasets are outlined in \citet[][]{schwope2022identification}; however, it should be mentioned that \citet[][]{littlefield2021quasi} points out that using \tess\ PDCSAP flux light curve data can suppress slow stochastic variability of true astrophysical origin. Therefore, here we reproduce the data from Sector 6 in the same manner as in \citet[][]{schwope2022identification}, but for Sector 33 and 87 we use the \tess\ light curve constructed from the \tess\ Simple Aperture Photometry (SAP) flux. All \tess\ data were downloaded using the \texttt{Lightkurve} package\footnote{\url{https://github.com/lightkurve/lightkurve}} \citep[][]{lightkurve2018ascl.soft12013L}, which was also used to remove the cosmic rays. All data points with quality flag greater than 0 were excluded.

The main strength of \tess\ comes from its exceptional relative photometry. However, in order to convert it to absolute photometry, other simultaneous ground-based data is required. For this purpose we use quasi-simultaneous \asassn\ $g$-band data \citep[][]{Shappee2014, kochanek_2017} similarly to \citet[][]{Scaringi2012a} and \citet[][]{Veresvarska2024}. With the $g$-band central wavelength of 475~nm and 140~nm width, the band is partially overlapping with \tess\ passband (600 -- 1000 nm). Assuming no colour-term variations, the data obtained from the \asassn\ webpage can be used to convert the \tess\ flux in $\rm e^{-}s^{-1}$ to \asassn\ $g$-band flux in mJy. We do this by adopting the following linear relation between the fluxes $F_{\rm ASAS-SN} \left[ \rm mJy \right] = A \times F_{ \rm TESS} \left[ \rm e^{-}s^{-1} \right]  + C$. The fully converted \tess\ data with the used \asassn\ $g$-band light curve is shown in Figure \ref{fig:2SXPS_J0623_TESS_lightcurves}. The dashed grey line in the middle panel showing \tess\ Sector 33 data denotes a cut-off point, after which the light curve is no longer stationary. As such, the data beyond BJD-245700 $\sim$ 2225 is not used in this work. 

\begin{figure}
        \centering
        \includegraphics[width = \columnwidth]{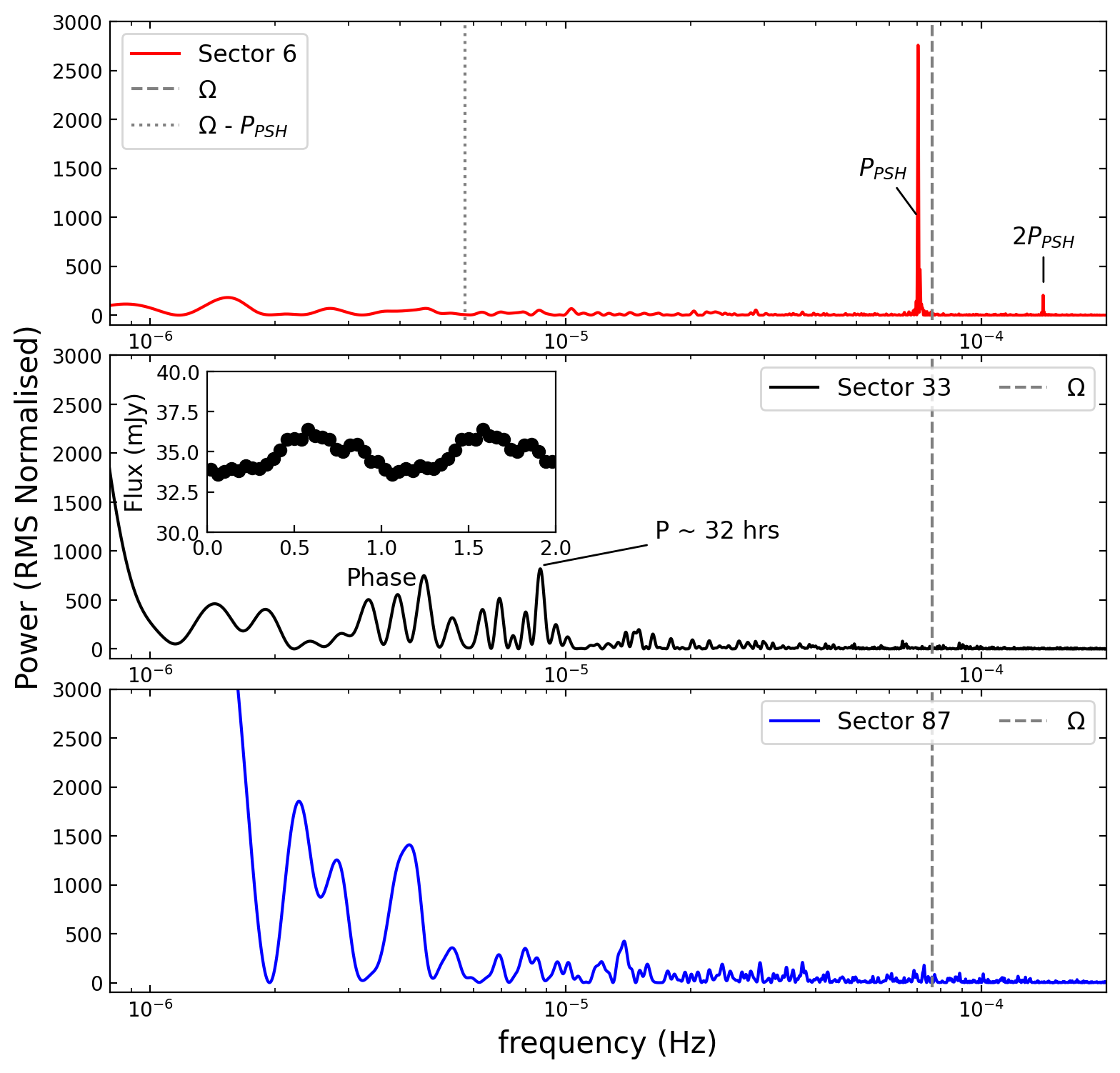}
        \caption[\textit{TESS} power spectra]{\textit{TESS} Sector 6 (top), Sector 33 (middle) and Sector 87 (bottom) power spectra. The 3.941-hour period ($P_{PSH}$) is indicated in the Sector 6 plot together with the first harmonic ($2P_{PSH}$). The Sector 33 plot indicates the low frequency signal at $\sim$32 hours, with an inset plot showing the phase-folded light curve on this period. The dashed line indicates the spectroscopic period of 3.645 hours associated with the orbital period.}
        \label{fig:2SXPS_J0623_TESS_power_spectra}
\end{figure}

In order to examine any potential periodic signals in \tess\ data, we compute the Lomb-Scargle periodogram of each separate Sector. The zoom in of this is shown in Figure \ref{fig:2SXPS_J0623_TESS_power_spectra}. As reported in \citet[][]{schwope2022identification}, Sector 6 shows a distinct coherent period at 3.941 hours and its first harmonic (marked as $P_{PSH}$ and $2P_{PSH}$ in Figure \ref{fig:2SXPS_J0623_TESS_power_spectra}). Both Sectors 33 and 87 show no evidence of this signal. Sector 33 however shows a potential signal at $\sim$32 hours. The phase-folded light curve on this period, binned into 50 phase bins, is shown in the inset plot of the middle panel of Figure \ref{fig:2SXPS_J0623_TESS_power_spectra}. The newest Sector 87 shows no sign of any periodic signals. 

\begin{figure}
        \centering
        \includegraphics[width = \columnwidth]{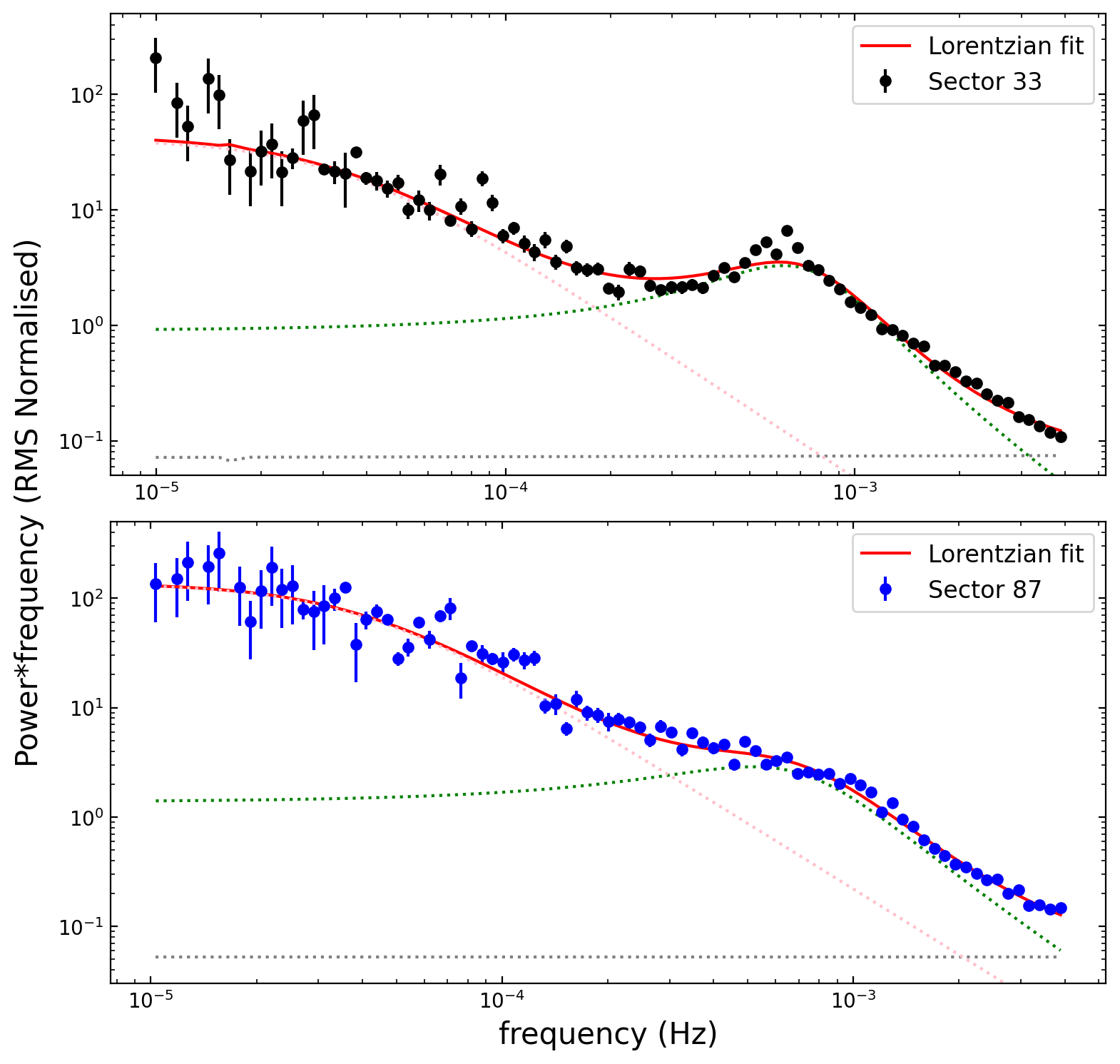}
        \caption[\textit{TESS} time-averaged power spectra]{\textit{TESS} Sector 33 (top) and Sector 87 (bottom) time-averaged power spectra. Both panels include the Lorentzian fit (solid red line) with the individual components shown in dotted lines.}
        \label{fig:2SXPS_J0623_TESS_TPS}
\end{figure}

A time-averaged power spectrum (TPS) is computed for Sectors 33 and 87 of \tess\ data and shown in Figure \ref{fig:2SXPS_J0623_TESS_TPS}. The TPS is constructed by separating the light curve of an individual Sector into 4-day segments. A segment of 4 days is chosen to maximise the number of segments in each Sector whilst minimizing data gaps in each segment. A Lomb-Scargle periodogram of each of the segments is then computed, before the individual power spectral densities (PSDs) are averaged together and binned onto a coarser frequency bin ($N_{\rm bins}=90$). The TPS allows for an easier examination of the broad-band shape of the power spectrum and increased signal-to-noise ratio that facilitates fitting the broad-band components of the TPS. To fit the TPS, we use Lorentzians and a Poisson noise component, similarly to \cite{Veresvarska2024MNRAS.534.3087V}:
                        \begin{eqnarray*}
    P_{\nu} = \sum_{i=0}^{N=2} P_{L} \left( r_{i}, \Delta_{i}, \nu_{0,i}, \nu \right) +  A,
\label{eq:PSD}
                        \end{eqnarray*}
\noindent where $A$ represents the Poisson noise level and $P_{L} \left( \nu \right)$ stands for the RMS normalized power of each individual Lorentzian, such that
                        \begin{eqnarray*}
    P_{L} \left( \nu \right) = \frac{r^{2} \Delta}{\pi} \frac{1}{\Delta^{2} + \left( \nu - \nu_{0} \right)^{2}},
\label{eq:Lorentz}
                        \end{eqnarray*}
\noindent with $\Delta$ corresponding to the half width half maximum (HWHM) and $r$ a normalization factor proportional to the integrated fractional rms. $\nu_{0}$ is the centroid frequency of the Lorentzian, with the frequency of the peak of the Lorentzian being defined as $\nu_{max} = \sqrt{\nu_{0}^{2} + \Delta^{2}}$ for $\nu_{0} > 0$. In the case of the low-frequency Lorentzian, it is assumed to be zero-centred ($\nu_{0,2} = 0$), as in the preliminary fitting $\nu_{0,2}$ is consistent with 0. Furthermore, the limiting low frequency of the TPS ($\sim10^{-5}$~Hz) does not allow for the detection of a defined structure of the broad-band feature to the same precision as in high frequencies and hence an extra free parameter in the form of $\nu_{0,2}$ is considered obsolete. Similarly to \cite{Veresvarska2024MNRAS.534.3087V}, whereas the low-frequency zero-centred Lorentzian can be fitted, it only represents an upper limit on its intrinsic value.

We perform a fit using the affine-invariant ensemble sampler
for Markov chain Monte Carlo (MCMC) as implemented by \texttt{emcee} \citep{emcee2013PASP..125..306F}. We have assumed priors with a small uniform perturbation centred on the best fit values obtained using Levenberg-Marquardt least-squares fitting method. The best fit values with their corresponding errors and reduced $\chi^{2}$ are detailed in Table \ref{tab:PSDfit}. The errors correspond to the confidence contours of 99.7$\%$ (3~$\sigma$). 
The effective peak frequency of the high frequency Lorentzian with corresponding error is also shown in Table \ref{tab:PSDfit}.

\begin{table}[h!]
\centering 
\caption{Values of the best fit parameters for the TPS in Figure \ref{fig:2SXPS_J0623_TESS_TPS} with corresponding errors.} 
\label{tab:PSDfit}
\begin{threeparttable}
\begin{tabular}{lcc}
\hline
\hline
    & Sector 33    & Sector 87   \\
    \hline
	$A$ $(\times 10^{-2})$ & 7$^{+1}_{-2}$ & 5$^{+2}_{-2}$   \\   
	$r_{1}$ $(\times 10^{-2})$ & 6.3$^{+0.2}_{-0.2}$  & 6.7$^{+0.6}_{-0.7}$ \\  
	$\Delta_{1}$ $(\times 10^{-4})$ & 3.8$^{+0.5}_{-0.5}$ & 4.9$^{+0.6}_{-0.5}$ \\ 
	$\nu_{0,1}$ $(\times 10^{-4})$ & 6.2$^{+0.3}_{-0.3}$  & 5$^{+1}_{-1}$ \\   
	$r_{2}$ $(\times 10^{-2})$ & 6.6$^{+0.4}_{-0.4}$  & 13.1$^{+0.8}_{-0.8}$	\\   
	$\Delta_{2}$ $(\times 10^{-5})$ & 3.4$^{+0.9}_{-0.8}$  & 4$^{+1}_{-1}$ \\    
    \hline
    $\chi^{2}_{\nu}$     & 3.2 & 3.5 \\
    $\nu_{max,1}$ $(\times 10^{-4})$ & 7.3$^{+0.6}_{-0.5}$  & 7$^{+1}_{-1}$ \\  
	\hline
	\hline
\end{tabular}
\end{threeparttable}
\end{table}

\subsection{Archival photometry}

Archival $V$-band photometry of SRGt 062340 from the Catalina Real-Time Transients Survey \cite[CRTS, ][]{drake2009first} was obtained between 2005 August 28 and 2013 May 9 and consists of 196 observations. Figure \ref{fig:2SXPS_J0623_archival_photometry} show the light curve from the CRTS data in red. There is clear variability, typically $\pm \sim$0.4 mag for most of the coverage, although there are two dips seen, at HJD 2454065 and HJD 2456362, where the magnitude of the system drops dramatically. Following a gap in the data (HJD 2456042 $-$ HJD 2456294) the object was much fainter than before (by $>$0.5 mag), reaching a minimum at HJD 2456362 at $V$ = 13.81. 

Archival ASAS-SN \citep{kochanek_2017} \footnote{https://asas-sn.osu.edu/} differential photometry was found (ASAS-SN Sky Patrol ID: 584115992288), spanning from 2013 October 30 to 2025 May 19, consisting of 6292 observations.
The light curve is shown in Figure \ref{fig:2SXPS_J0623_archival_photometry}, in blue and green data points, and where the spectroscopic and photometric observations reported on in this study is superimposed. Consult appendix \ref{App:Obs_logs} and text for more detail on the observation epochs, as there is significant overlap for some observations, making it difficult to distinguish between them in the light curve. There is variability, of amplitude $\lesssim$ 0.5 mag, for the majority of the ASAS-AN light curve, with a slight dip and then brightening between JD $\sim$ 2458000 - 2459000. However, a significant and sudden decrease in brightness, $\gtrsim$ 3 mag, is seen from JD $\sim$ 2460200. A minimum is reached at JD $\sim$ 2460400, after which it re-brightens, again with a similarly steep rate as the dimming. There is another dip shortly after; however, it is not as sudden as the first. The system appear to still be in a dip at the last data points. These drops in brightness is reminiscent of the VY Scl class of CV.       

\section{Discussion} \label{Sec:Discussion}

The spectroscopic observations obtained with the SAAO 1.9m telescope in 2021 February and March showed similar broad Balmer absorption lines, whose centres are filled in by prominent narrow emission lines, as was observed with the ANU/WiFeS spectrograph on 2020 October 28 (HJD 2459150.6640), while the \ion{He}{ii} $\lambda$4686 and Bowen fluorescence lines were also seen, as was observed with SALT on October 31 (HJD 2459154.4672) \citep[][]{schwope2022identification}, albeit at much lower resolution.
These Balmer spectral features are similar to what \citet[][]{stickland1984rz} observed for the NL CV RZ Gru, even though their study focused more into the far blue and UV, in which they attributed the absorption features to a thick accretion disc, and the narrow emission lines being as a result of the low inclination angle, $i < 20^{\circ}$, RZ Gru is observed at.  The similarities between the spectra of RZ Gru and SRGt 062340 could possibly indicate that it is a UX UMa system seen at a low inclination. UX UMa's are sometimes referred to as `thick-disc CVs' based on their absorption lines \citep[][]{warner2003cataclysmic}, and are a subclass of the NL CVs. \cite{Skillman1995PASP..107..545S} also found similarly low K-amplitude (25$\pm$4 km s$^{-1}$) for the emission line radial velocities of MV Lyrae, and concluded that the system is seen at a low inclination (10$^\circ$ - 13$^\circ$). Even though the spectrum that is shown by \citet[][]{Skillman1995PASP..107..545S} (Figure 1 in their paper) extends further to the red than Figure \ref{fig:SGRt_0623_spectrum}, it is still very similar to what is seen for SRGt 062340. Given this, and the fact that MV Lyrae has a similar orbital period ($\sim$3.2 hours) as SRGt 062340, further supports our interpretation that SRGt 062340 is likely seen at a very low inclination.

The H$\beta$ and H$\gamma$ spectral lines are also similar to those observed in ASAS J071404+7004.3, another bright and nearby NL that has only recently been identified as a CV \citep[][]{inight2022asas}, where they also  attribute an optically thick accretion disc responsible for the Balmer absorption lines. They propose a disc wind model to explain the strong single-peaked emission lines. 

Both optically thick and thin discs, if observed at high inclinations, usually display double peaked emission lines \citep[][]{horne1986emission}. However, the presence of a disc wind can significantly alter the profile of emission lines, resulting in high-inclination systems to also show single peak emission lines \citep[][]{Honeycutt_disk_wind, Matthews_disk_wind, Tampo_disk_wind}. 
The SALT HRS observations revealed a non-Gaussian profile for the H$\alpha$ emission line, where the strength, primarily of the red peak, is seen to vary slightly on timescales as short as tens of minutes, as revealed in Figure \ref{fig:2SXPS_J0623_HRS_stacked_spectra_example}. In addition, trailed spectra (Figures \ref{fig:SGRt_0623_trailed_spectra} and \ref{fig:SGRt_0623_phase_folded_trail_spectra}) and photometry show no indication of any kind of eclipse by the secondary star, from which it can be concluded that the inclination of the system must be low enough for the secondary to not eclipse the region where H$\beta$ or H$\gamma$ emission is produced, implying $i \lesssim 75^{\circ}$, and possibly even much smaller given the low amplitude variations in the radial velocity measurements of the emission lines. The decisive feature in the spectrum of ASAS J071704+7004.3 that strongly indicate the presence of a disc wind is the P Cygni profile, seen for \ion{He}{i} $\lambda$5876. There is slight indication of a possible transient P Cygni profile to the \ion{He}{i} $\lambda$5876, line seen from the SALT HRS observations. However, the normalization of the HRS spectra makes it difficult to see any absorption components, even for H$\beta$ and H$\gamma$, where we know from the SpUpNIC observations there is absorption.

The Lomb-Scargle period analysis of $\sim$52 hours of high-speed SAAO photometry, taken on 18 nights over a time-base of $\sim$4 months, has revealed evidence of a weak (few percent amplitude) period at 24.905 min (Figure \ref{fig:2SXPS_J0623_photometry_power_spectrum}).
While some individual nights show strong period peaks in this region (Figure \ref{fig:2SXPS_J0623_lightcurve_set} in the appendix), they are not always present, which is likely a consequence of the extreme stochastic flickering. Figure \ref{fig:2SXPS_J0623_lightcurve_set} also reveals that numerous flaring and dipping features appear in the light curves which, at first glance, appear identical in shape and duration and therefore might be caused by the same mechanism, but judging from our power spectra these are incoherent variations, so not orbital or spin phase dependant. This stochastic non-periodic, and often high amplitude, variability is characteristic of flickering in many magnetic CVs, and particularly in some IPs  \citep[e.g. TX Col, ][]{1989ApJ...344..376B}. In fact, if the 24.905-min period found from the SAAO photometric measurements is taken to be the spin period, this will give a spin-to-orbital period ratio, P$_\text{{spin}}$/P$_\text{{orb}}$, of 0.114. Such ratios are typical of IPs \citep[][]{barrett1988photometry, mondal2022intermediate}. 

Assuming that the orbital period from spectroscopy is 3.645 hours (Section \ref{sss:spec_orb}, and in agreement with the period found from X-ray analysis by \citet[][]{cuneo2026A&A...705A..71C}), the signal in \tess\ Sector 6 could be interpreted as a positive superhump. In the case of positive superhumps, $\frac{1}{P_{\rm PSH}} < \Omega$, and the signal is associated with the tidal stresses exerted by the secondary star on the accretion disc. These cause the disc to become eccentric and undergo apsidal precession in the prograde direction \citep{Lubow1991}. If this is the case, then we would expect the fundamental precession frequency at $\frac{1}{P_{+}} = \Omega - \frac{1}{P_{\rm PSH}} \sim$ 48 hours. However, no low frequency signal is detected in \tess\ Sector 6. This may be due to the data quality, as Sector 6 has a low cadence of $\sim$ 30 minutes and the scatter is larger than in Sector 33, where a low frequency periodicity can also be seen by eye in Figure \ref{fig:2SXPS_J0623_TESS_lightcurves}. This unfortunately means that we cannot unambiguously confirm the nature of the signal from Sector 6. 

Assuming the signal from Sector 6 is a positive superhump, we can use it to compute the excess $\epsilon = \frac{1/P_{PSH} - \Omega }{\Omega} \sim $~0.081. Using the relation observed between the superhump excess and orbital period \citep{Patterson1998PASP..110.1132P} we can show that SRGt 062340 lies on this relation, adding more evidence towards the positive superhump hypothesis. \cite{2005PASP..117.1204P} extends this relations to higher values of q, with BB Dor $\epsilon = 0.0939(15)$. Equation 8 from \cite{2005PASP..117.1204P} also allows us to estimate a value of the binary mass ratio q, such that
\begin{eqnarray*}
    \epsilon = 0.18 \rm q + 0.29 \rm q^{2},
\label{eq:eps}
\end{eqnarray*}
from which we obtain q $\sim$ 0.30. This would suggest $M_{2} \sim$ 0.24 $M_{\odot}$, assuming that $M_{\rm WD} \sim$ 0.8 $M_{\odot}$. This would then place SRGt 062340 in Figure 9 from \cite{2005PASP..117.1204P} between BB Dor and DW UMa and UU Aqr, in accordance with the observed relation between $\Omega$ and $\epsilon$.

If SRGt 062340 is indeed a NL VY Scl system which is compatible with the long-term \asassn\ light curve, a presence of positive superhumps is unusual. This is because the large disc required for positive superhumps is more easily attained in systems with $P_{orb} <$ 2 hrs due to the compact nature of the systems. However, there are some systems showing photometrically similar behaviour as SRGt 062340. In particular, BB Dor, a NL of similar orbital period to SRGt 062340. In particular, the positive superhump seen in BB Dor is also associated with a similar signal in low frequencies as seen in SRGt 062340 in Sector 33. Another similar NL is DW UMa, which shows only positive superhumps at similar orbital period to SRGt 062340 \citep{Bruch2023MNRAS.519..352B}. MV Lyr is another NL at similar orbital period, and also very low orbital inclination \citep{Skillman1995PASP..107..545S}, resulting in the orbital period not being detectable in \tess\,.

In summary, it is likely that the signal in Sector 6 is a positive superhump, but it cannot be proven with the data at hand. Furthermore, it may be possible that once the system returns to high state the superhump will reappear, allowing a confirmation of this hypothesis.

Flickering, or aperiodic broad-band variability, occurs in all accreting systems \citep{Scaringi2012,Uttley2001,Uttley2005,VandeSande2015,Gandhi2009}. Observations confirm this behaviour \citep{Belloni2002,McHardy2006,Scaringi2012a}, and it is widely attributed that some flickering could be due to local accretion rate fluctuations propagating inwards on viscous timescales \citep{Lyubarskii1997, AU06}. Material moves inwards as angular momentum is transferred outwards due to viscous stresses between disc rings, which rotate at different Keplerian velocities \citep{Frank2002}. The standard Shakura-Sunyaev disc model \citep{Shakura1973} introduces a dimensionless viscosity parameter, $\alpha$, often assumed constant. However, flickering likely results from local viscosity fluctuations \citep{Lyubarskii1997}.

The fluctuating accretion disc model has been successfully applied to X-ray binaries, where it describes variability in the optically thin, geometrically thick inner flow (or corona) at X-ray wavelengths \citep{Klis2006}, and to CVs at optical wavelengths \citep{Scaringi2014b}. In these cases a break feature in the time-averaged power spectrum (TPS) of a light curve displaying flickering shows a `break' which in the framework of propagating viscosity fluctuations can be assumed to occur at viscous timescales, such that the associated frequency is
\begin{eqnarray*}
    \nu_{visc} = \alpha \left( \frac{H}{R} \right) ^{2} \sqrt{\frac{GM}{R^{3}}},
\end{eqnarray*}
\noindent where $\frac{H}{R}$ is the scale height ratio at distance R away from the centre of the accretor of mass $M$ and $G$ is the gravitational constant. Assuming that the periods of 24.907 and 28.630 minutes from Section \ref{subsec:saao_photometry} are associated with the spin, we can estimate the corresponding dynamical frequency at co-rotation radius $R_{CO}$
\begin{eqnarray*}
    R_{CO} = \left(  \frac{GMP_{\omega}^{2}}{4 \pi^{2}} \right)^{3},
\end{eqnarray*}
\noindent where $P_{\omega}$ is the spin period of the system. Assuming that $\nu_{\max}$ from Table \ref{tab:PSDfit} corresponds to the viscous frequency at the co-rotation radius, we can estimate $\alpha\left( \frac{H}{R} \right)^2$ under the assumption of a truncated accretion disc scenario found in IPs and suggested to be present in VY Scl systems by \citet[][]{VYScl_msgnetic}. Given the negligible difference in $\nu_{\max}$ between Sectors 33 and 87 and the two potential spin periods, we find $\alpha\left( \frac{H}{R} \right)^2 \sim 0.17 - 0.2$. This is slightly lower than the value of $\sim 0.7$ obtained from modelling the TPS of MV Lyr in \cite{Scaringi2014b}. This discrepancy aligns with the interpretation of SRGt 062340 as an IP with a truncated accretion disc, in contrast to MV Lyr, which is classified as a standard VY Scl-like system with traces of an inner, hot, geometrically thick, and optically thin flow. Furthermore, our results are consistent with the expectation that $\alpha\left( \frac{H}{R} \right)^2$ decreases with increasing radius \citep{Veresvarska2023}. In order for the break to occur at the WD radius $R_{WD} \sim 0.0099 R_{\odot}$, the $\alpha\left( \frac{H}{R} \right)^2 \sim 0.001$, much smaller than the estimated MV Lyr value for inner disc. Whereas this cannot confirm the IP nature of SRGt 062340 or unambiguously determine the spin period, it can serve as circumstantial evidence suggesting that the broad-band structure of the TPS is consistent with an IP-like truncated disc. However, without a firm detection of a spin period, this interpretation remains purely speculative.

The long-term light curves from CRTS and ASAS-SN, Figure \ref{fig:2SXPS_J0623_archival_photometry}, reveal long-term low amplitude and slow modulation in brightness, of $\sim$0.2 mag as well as dips, $\sim$1.5 mag in CRTS and $\sim$4 mag in ASAS-SN, which last at least several weeks. 
VY Scl systems are defined as nova-likes showing a drop in luminosity of mag > 1 \citep[][]{King_VY_Scl}, suggesting that SRGt 062340 belongs to this subgroup. However, these periodic dips in luminosity, together with the low amplitude and slow modulation in the brightness, is consistent with what has also been observed for IPs \citep[][]{warner2003cataclysmic, Covington_low_flux} and therefore an IP interpretation is also plausible. However, because of the lack of a clear and consistent spin period, the IP interpretation cannot be either confirmed or refuted. Polarimetric observations were conducted at the SAAO; however, no significant evidence of polarization was detected (Buckley, private communication). This lack of polarization also cannot definitively rule out the possible IP nature of the system.

\begin{figure}
        \centering
        \includegraphics[width = \columnwidth]{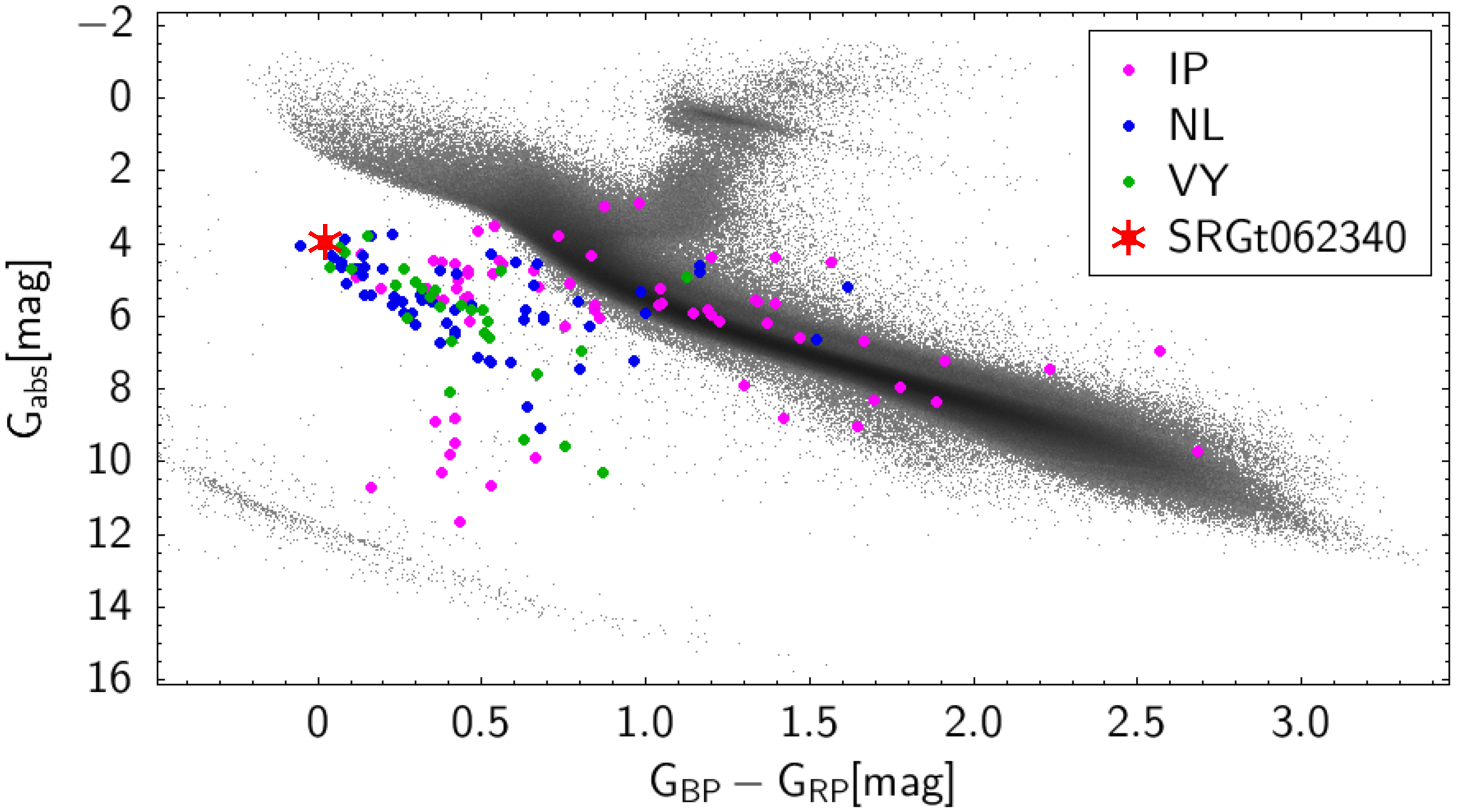}
        \caption[Colour Map]{Gaia colour-magnitude diagram showing the positions of NLs (blue), VY Scl stars (green) and IPs (magenta), with the location of SRGt 062340 indicated by the red star.}
        \label{fig:2SXPS_J0623_CM}
\end{figure}

Figure \ref{fig:2SXPS_J0623_CM} show colour-magnitude plots of NL systems, including the VY Scl subclass and IPs, where the colour and distance parameters were obtained from Gaia EDR3. Using the Gaia colour and absolute magnitude clearly indicates that SRGt 062340 is located at the expected region to be a NL system \citep[][]{Dubus_Gaia}. Based on its position in this colour-magnitude diagram, it appears that SRGt 062340 is more likely to be a VY Scl system than an IP; however, SRGt 062340 seems to be slightly more blue than any VY Scl system in this plot.

Another phenomenon that needs to be considered in interpreting the power spectra of the SAAO photometry is the possible presence of quasi-periodic oscillations, QPOs (\cite{patterson1977rapid}). Although multiple peaks are seen in the power spectrum of the SAAO photometric observations in the frequency range expected of QPOs (Figure \ref{fig:2SXPS_J0623_photometry_power_spectrum}), it is not immediately clear whether QPOs are present, since they usually appear as broad and noisy bumps in the power spectra. 
The changing periodicities found between each individual night's observations, however, could suggest their presence, although in some IPs, for example TX Col (\cite{buckley1992remarkable}, \cite{mhlahlo2007discovery} and \cite{littlefield2021quasi}), the purported WD spin period is not always seen to be overwhelmingly prominent, and at times even absent, in period analysis. This is likely due to strong stochastic flickering and the possible presence of red noise. Therefore, the WD spin period might be overwhelmed by the presence of QPO's on a single night's observation; however, using all the photometric observations, with a long baseline, the WD spin period is more likely to be detected.   

\section{Conclusions} \label{Sec:Conclusion}

For this NL CV, a possible WD spin period of P$_{\text{spin}}$ = 24.905 min was found from an analysis of high-speed photometric observations, while a P$_{\text{orb}}$ = 3.654-hour period was found from time-resolved spectroscopic observations by analysing the H$\beta$ and H$\gamma$ emission line radial velocities. The probable spin-orbital period ratio is P$_{\text{spin}}$/P$_{\text{orb}}$ = 0.114, which places it away from confirmed IPs with orbital periods above the period gap \citep[see Figure 11 in][]{Mukai_X_ray}.
The presence of deep dips in brightness seen in long-term light curves (Figure  \ref{fig:2SXPS_J0623_archival_photometry}), and its location in the colour-magnitude plot (Figure \ref{fig:2SXPS_J0623_CM}), also suggests that SRGt 062340 is more likely a VY Scl system. 

The radial velocities of the H$\beta$ and H$\gamma$ emission lines were thoroughly analysed in this study; however, wider wavelength coverage of time-resolved spectroscopic observations that include longer wavelength such as the H$\alpha$ line could help to confirm the orbital period of the system. SALT HRS observations were obtained, which included H$\alpha$; however, these observations were taken to study the morphology of the emission lines, and showed that, at least for H$\alpha$, the lines show slight indication of having a non-Gaussian morphology, and of having a high-velocity component in some spectra. These observations also revealed that the morphology of the H$\alpha$ emission line does seem to change; however, longer coverage per night is needed to determine whether these changes are phase related. No sign of emission from the irradiated face of the secondary star was however seen.  
In addition, UV observations (e.g. with HST) could reveal disc wind features (such as P-Cygni profiles) often more easily seen at shorter wavelengths.

\begin{acknowledgements}
      JB acknowledges support from the
      \emph{Deut\-sche For\-schungs\-ge\-mein\-schaft, DFG\/} project
      number Ts~17/2--1, and the South African National Research Foundation (NRF).
      Some of the observations reported in this paper were obtained with the Southern African Large Telescope (SALT) under program 2021-2-LSP-001 (PI: D.\ Buckley). Polish participation in SALT is funded by grant No.\ MEiN nr 2021/WK/01. 
      We thank the South African Astronomical Observatory for generous allocations of observing time.
      MV acknowledges the support of the Science and Technology Facilities Council (STFC) studentship ST/W507428/1.
      VAC acknowledges support from the Deutsches Zentrum für Luft- und Raumfahrt (DLR) under contract number 50 OR 2405
\end{acknowledgements}

\bibliographystyle{aa}
\bibliography{bibliography}

\begin{appendix} 

    \section{Observing logs} \label{App:Obs_logs}
    
    \begin{table}[h!]
    \centering 
    \caption{SAAO 1.9m SpUpNIC spectroscopic observations} 
    \label{tab:2SXPS_obs_log}
    \begin{tabular}{ccccccc}
    \hline
    Date  		& HJD  			& N  	& Exposure Time	    & Duration & Grating & Wavelength   \\
     			& {(start)} & (observations) & (seconds) 	& (minutes) & Angle  & Coverage (\AA) \\ \hline

	2021-02-02  & 2459248.2966 & 9  &   1200 & 192.65    &   5.50    &  3795 - 5025   \\
	2021-02-26  & 2459272.2721 & 54 &   300  & 299.97    &   4.20    &  4116 - 5326   \\
	2021-02-27  & 2459273.2577 & 60 &   300  & 322.57    &   4.20    &  4116 - 5326   \\
	2021-03-02  & 2459276.2542 & 54 &   300  & 289.63    &   4.20    &  4116 - 5326   \\
	2021-03-17  & 2459291.2892 & 15 &   300  & 156.55    &   5.50    &  3762 - 4990   \\
	2021-03-18  & 2459292.2858 & 24 &   300  & 155.05    &   5.50    &  3760 - 4992   \\
	2021-03-19  & 2459293.2916 & 20 &   300  & 128.50    &   5.50    &  3768 - 5000   \\
	2021-03-20  & 2459294.2814 & 10 &   300  & 62.68     &   5.50    &  3768 - 5000   \\
	2021-03-21  & 2459295.3110 & 20 &   300  & 123.10    &   5.50    &  3800 - 5029   \\
	2021-03-22  & 2459296.2745 & 28 &   300  & 170.85    &   5.50    &  3800 - 5029   \\
	\hline
	\hline
    \end{tabular}
    \end{table}

    \begin{table}[h!]
    \centering 
    \caption{SAAO photometric observation log} 
    \label{tab:2SXPS_photometry}
    \begin{threeparttable}
    \begin{tabular}{ccccccc}
    \hline
    Date  		& BJD  			& Exposure Time  		& Duration & Filter	& Telescope & SHOC   \\
     	  		& {(start)}     & (seconds) 	        & (hours)  &        &           & Camera \\ \hline

	2020-11-01	                &   2459155.5004	&   0.3      &   3.213   &   clear   & 40 inch   &   SHA \\
	2020-11-02\tnote{$^\star$}	&   2459156.5195    &   1.0      &   2.407   &   clear   & 40 inch   &   SHA \\
	2020-11-03\tnote{$^\star$}	&   2459157.5407    &   0.3      &   1.046   &   clear   & 40 inch   &   SHA \\
	2020-11-27	                &   2459181.4112    &   2.0      &   3.102   &   clear   & Lesedi    &   SHD \\
	2020-11-28	                &   2459182.3333    &   2.0      &   1.905   &   clear   & Lesedi    &   SHD \\
	2020-11-29	                &   2459183.3305    &   2.0      &   6.084   &   clear   & Lesedi    &   SHD \\
	2020-12-03\tnote{$^\star$}	&   2459187.5572    &   5.0      &   0.202   &   g'      & 40 inch   &   SHA \\
	2020-12-04\tnote{$^\star$}	&   2459188.4346    &   10.0     &   0.707   &  clear    & 40 inch   &   SHA \\
	2021-01-04	                &   2459219.5329    &   1.0      &   1.580   &  clear    & 40 inch   &   SHA \\
	2021-01-13	                &   2459228.3549    &   1.0      &   5.682   &  clear    & 40 inch   &   SHA \\	
	2021-01-20	                &   2459235.3732    &   1.0      &   5.065   &  clear    & 40 inch   &   SHA \\
	2021-02-01	                &   2459247.4396    &   1.0      &   1.244   &  clear    & Lesedi    &   SHD \\
	2021-02-02	                &   2459248.2718    &   1.0      &   1.419   &  clear    & Lesedi    &   SHD \\	
	2021-02-26	                &   2459272.2984    &   1.0      &   4.146   &  clear    & Lesedi    &   SHD \\
	2021-02-27	                &   2459273.3171    &   1.0      &   4.216   &  g'       & 40 inch   &   SHA \\
	2021-02-27                  &   2459273.2633    &   1.0      &   4.919   &  clear    & Lesedi    &   SHD \\
	2021-03-02	           	    &   2459276.3000    &   5.0      &   4.485   &  g'       & 40 inch   &   SHA \\
	2021-03-02                  &   2459276.2462    &   1.0      &   5.092   &  r'       & Lesedi    &   SHD \\
	\hline
	\hline
    \end{tabular}
    \begin{tablenotes}
    \item[$^\star$] Observation of poor quality and was not used in analysis 
    \end{tablenotes}
    \end{threeparttable}
    \end{table}

    \begin{table}[h!]
    \centering 
    \caption{SALT HRS observations} 
    \label{tab:2SXPS_HRS_obs_log}
    \begin{tabular}{ccccccc}
    \hline
    Date  		& JD  			& N  	& Exposure Time	    & Duration  & Wavelength   \\
     			& {(start)} & (observations) & (seconds) 	& (minutes) & Coverage (\AA) \\ \hline

	2023-11-09  & 2460258.4463 & 4  &   600  & 43.5      &  3734 - 8789   \\
	2023-11-28  & 2460277.3959 & 2  &   600  & 21.2      &  3764 - 8789   \\	
	2023-11-29  & 2460278.3915 & 2  &   600  & 21.2      &  3764 - 8789   \\
	2023-11-30  & 2460279.3943 & 2  &   600  & 21.2      &  3764 - 8789   \\
	2023-12-07  & 2460286.3732 & 2  &   600  & 21.2      &  3764 - 8789   \\	 
	2023-12-14  & 2460293.3488 & 4  &   600  & 43.5      &  3764 - 8789   \\
	2024-01-13  & 2460323.4995 & 4  &   600  & 43.5      &  3764 - 8789   \\
	2024-01-15  & 2460325.4968 & 4  &   600  & 43.5      &  3764 - 8789   \\
	2024-01-20  & 2460330.4830 & 4  &   600  & 21.2      &  3764 - 8789   \\	
    2024-01-31  & 2460341.4521 & 4  &   600  & 43.5      &  3764 - 8789   \\
    2024-03-19  & 2460389.3187 & 4  &   600  & 43.5      &  3764 - 8789   \\
    2024-03-29  & 2460399.2967 & 4  &   600  & 43.5      &  3764 - 8789   \\
    2024-04-02  & 2460403.2837 & 4  &   600  & 43.5      &  3764 - 8789   \\
    2024-04-04  & 2460405.2806 & 4  &   600  & 43.5      &  3764 - 8789   \\
	\hline
	\hline
    \end{tabular}
    \end{table}

    \clearpage

    \section{Spectroscopy} \label{App:Spectroscopy}

    \begin{figure}[h!]
	\includegraphics[width=\columnwidth]{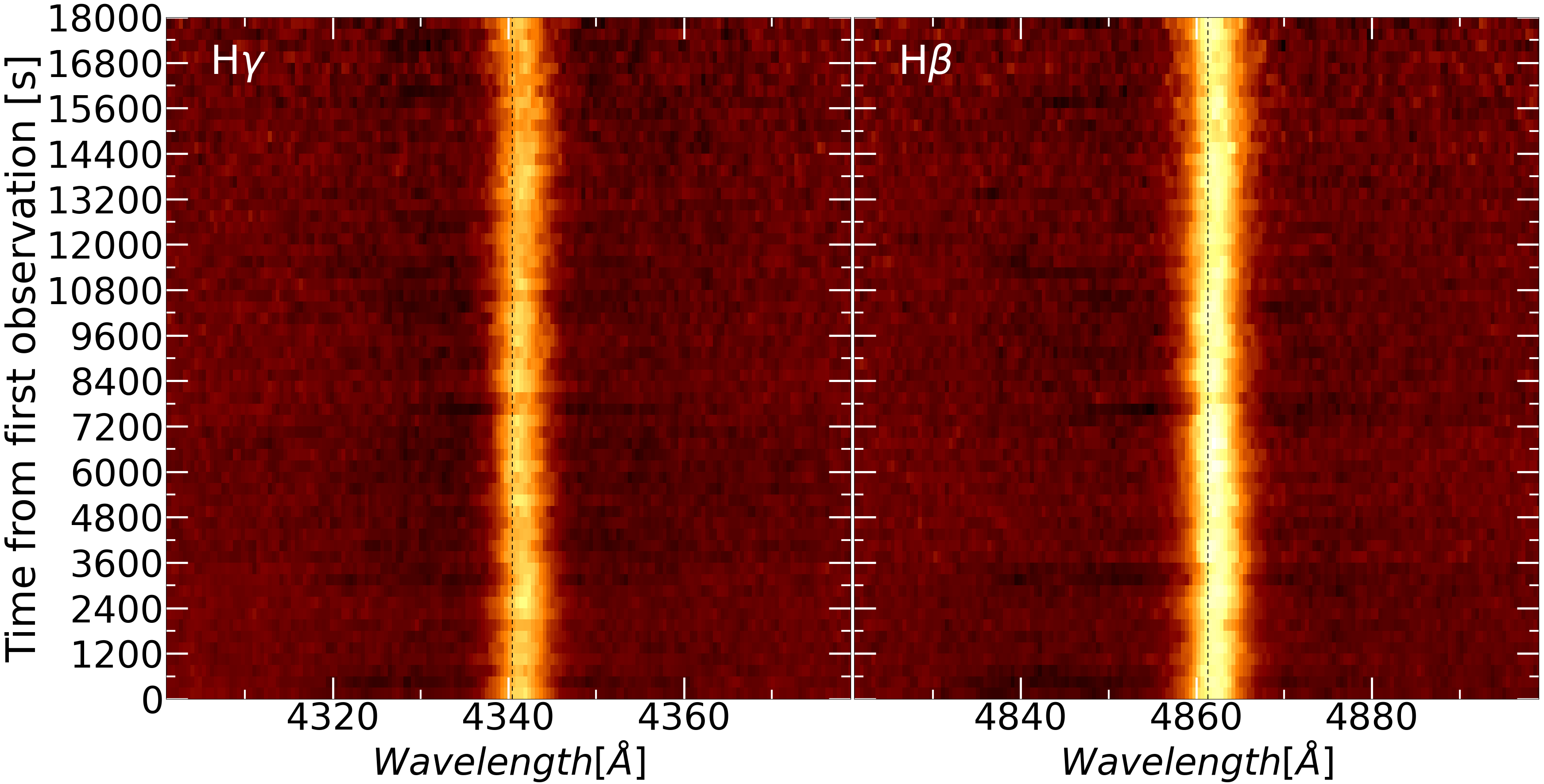}
        \caption[Trailed Spectra]{Trailed spectra of H$\gamma$ (left) and H$\beta$ (right) from 2021 February 27. The vertical dashed black lines coincide with the respective rest wavelengths, which show that the peak of the emission lines lie slightly towards longer wavelengths. Note the wider darker regions of the absorption lines, adjacent to both sides of the bright emission lines, and the sporadic stronger absorption.}
        \label{fig:SGRt_0623_trailed_spectra}
    \end{figure}

    \begin{table}[h!]
    \caption{Averaged emission line width measurements} 
    \label{tab:SGRt_0623_emission_line_summary}
    \begin{tabular}{@{}c|ccc}
    \hline
    \hline
    \multicolumn{1}{}{} &
    \multicolumn{1}{c}{H$\beta$} &
    \multicolumn{1}{r}{\ion{He}{ii} $\lambda$4686} &
    \multicolumn{1}{c}{H$\gamma$}\\
    \hline

Date                    & \multicolumn{3}{c}{FWHM (km s$^{-1}$)}   \\
    \hline
    2021 February 2     &  337 $\pm$ 8  &  394 $\pm$ 28  & 366 $\pm$ 9   \\
	2021 February 26	&  318 $\pm$ 8  &  403 $\pm$ 26  & 321 $\pm$ 8   \\
	2021 February 27	&  313 $\pm$ 9  &  414 $\pm$ 26  & 324 $\pm$ 8   \\
	2021 March 2    	&  327 $\pm$ 8  &  403 $\pm$ 25  & 327 $\pm$ 7   \\
	2021 March 17    	&  317 $\pm$ 7  &  403 $\pm$ 24  & 307 $\pm$ 9   \\
	2021 March 18    	&  313 $\pm$ 11 &  394 $\pm$ 40  & 306 $\pm$ 11  \\
	2021 March 19    	&  317 $\pm$ 9  &  399 $\pm$ 23  & 308 $\pm$ 10  \\
	2021 March 20    	&  303 $\pm$ 7  &  394 $\pm$ 23  & 291 $\pm$ 8   \\
	2021 March 21    	&  310 $\pm$ 8  &  402 $\pm$ 25  & 315 $\pm$ 7   \\
	2021 March 22    	&  310 $\pm$ 9  &  404 $\pm$ 26  & 309 $\pm$ 9   \\
	\hline
	\hline
\end{tabular}
\end{table}

\begin{table}[h!]
\centering 
\caption{Averaged absorption line width measurements} 
\label{tab:SGRt_0623_absorption_line_summary}
\begin{tabular}{c|cc}
\hline
\hline
\multicolumn{1}{}{} &
\multicolumn{1}{c}{H$\beta$} &
\multicolumn{1}{c}{H$\gamma$}\\
\hline

Date                    & \multicolumn{2}{c}{FWHM (km s$^{-1}$)}   \\
    \hline
    2021 February 2     &  4356 $\pm$ 309   &     3295 $\pm$ 140   \\ 
	2021 February 26	&  4213 $\pm$ 234	&     2831 $\pm$ 207   \\
	2021 February 27	&  4133 $\pm$ 256	&     2747 $\pm$ 213   \\
	2021 March 2    	&  4230 $\pm$ 215	&     2989 $\pm$ 184   \\
	2021 March 17    	&  4099 $\pm$ 174	&     2791 $\pm$ 147    \\
	2021 March 18    	&  4180 $\pm$ 331	&     3112 $\pm$ 345    \\
	2021 March 19    	&  4089 $\pm$ 248	&     2773 $\pm$ 212    \\
	2021 March 20    	&  3927 $\pm$ 219	&     2692 $\pm$ 170    \\
	2021 March 21    	&  4218 $\pm$ 270	&     3389 $\pm$ 298    \\
	2021 March 22    	&  4405 $\pm$ 359	&     3243 $\pm$ 322    \\
	\hline
	\hline
\end{tabular}
\end{table}

\begin{figure*}[h!]
        \centering
        \includegraphics[width = 1.0\textwidth]{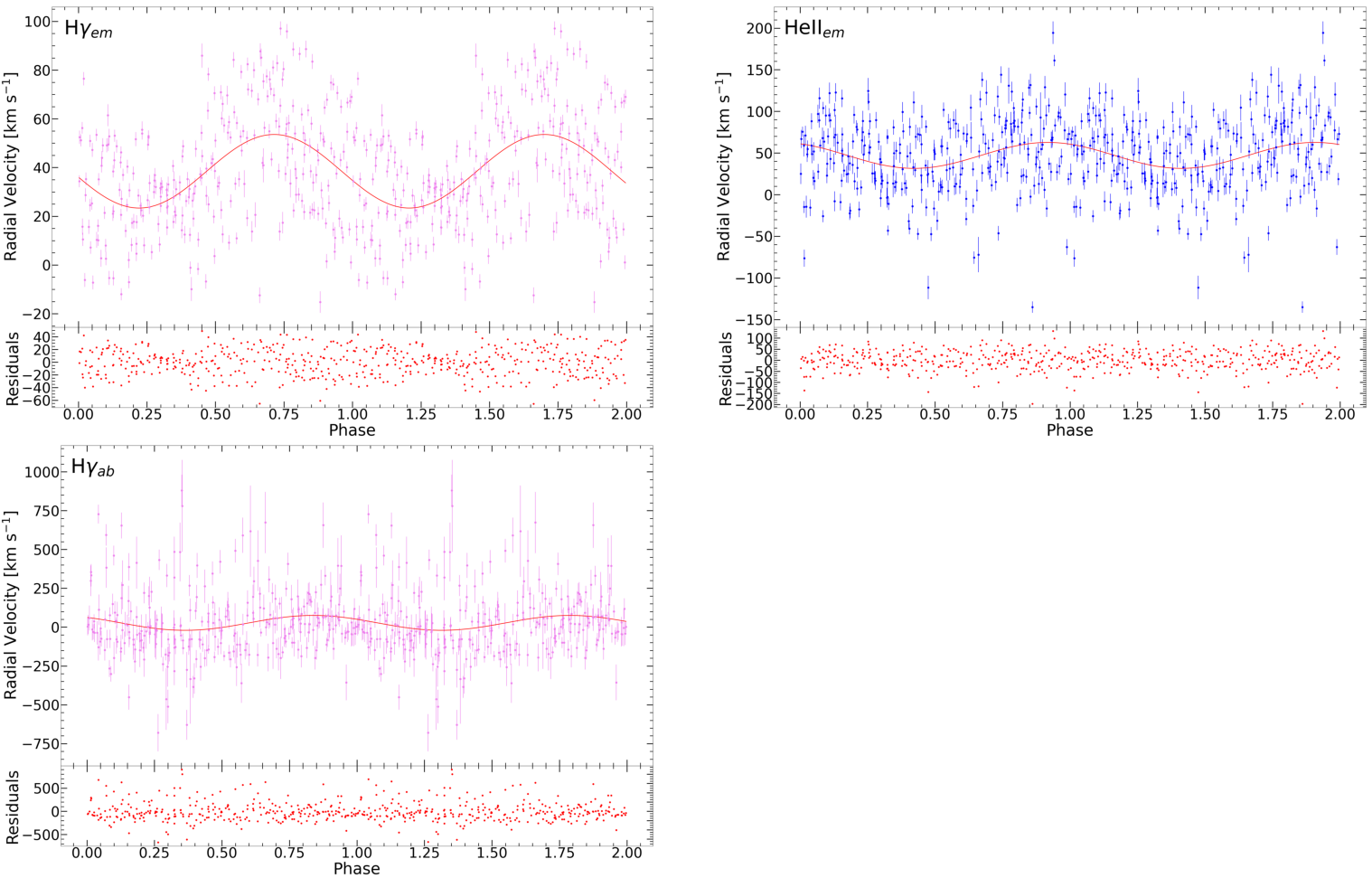}
        \caption[phase-folded radial velocities]{H$\gamma$ emission (H$\gamma_{em}$, top left) and absorption (H$\gamma_{ab}$, bottom left) line radial velocities phase-folded on the 3.645-hour period, with the phase-folded \ion{He}{ii} $\lambda$4686 emission line radial velocities at top right. The fitted sine functions, with the parameters in Table  \ref{tab:SGRt_0623_phase_folded_parametrs}, are also shown, with the residuals in the lower panels of each plot.}
        \label{fig:SGRt_0623_phase_folded_emission_absorption_rv}
\end{figure*}

\begin{table}[h!]
\centering 
\caption{SALT HRS radial velocity measurements} 
\label{tab:appendix_2SXPS_J0623_HRS_rv}
\begin{threeparttable}
\begin{tabular}{cccc}
\hline
\hline
HJD      &$\phi$    & H$\alpha$ Blue    & H$\alpha$ Red    \\
(mid)    &          & (km s$^{-1}$)     & (km s$^{-1}$)  \\
    \hline
    2460258.452078  &    0.102   &	-17.4   $\pm$	0.8	&	98.5	$\pm$	3.2	\\
    2460258.459823	&    0.161   &	-4.5	$\pm$	0.8	&	110.9	$\pm$	3.1	\\
    2460258.467589	&    0.220   &	1.9	    $\pm$	1.5	&	94.2	$\pm$	3.7	\\
    2460258.475343	&    0.278   &	5.4	    $\pm$	3.3	&	81.6	$\pm$	6.7	\\
	                &			 &             	        &				        \\
    2460277.402424	&    0.807   &	10.8	$\pm$	1.2	&	167.4	$\pm$	7.3	\\
    2460277.410190	&    0.866   &	-6.0	$\pm$	1.2	&	124.3	$\pm$	3.2	\\
	                &            &				        &				        \\
    2460278.398117	&    0.357   &	15.8	$\pm$	2.1	&	87.3	$\pm$	1.8	\\
    2460278.405872	&    0.416   &	-15.8	$\pm$	3.0	&	83.9	$\pm$	2.7 \\
	                &            &				        &		                \\
    2460279.400951	&    0.962   &	16.3	$\pm$	1.4	&	115.0	$\pm$	3.5	\\
    2460279.408705	&    0.021   &	23.5    $\pm$	0.6	&	170.8	$\pm$	2.2	\\
	                &            &				        &		                \\
    2460286.379980	&    0.886   & 	-21.0   $\pm$	1.0	&	137.1	$\pm$	4.3	\\
    2460286.387734	&    0.944   &	-24.9   $\pm$	0.9	&	136.4	$\pm$	3.2	\\
	                &            &				        &		                \\
    2460293.355776	&    0.785   &	-18.9   $\pm$	0.9	&	119.0	$\pm$	4.3	\\
    2460293.363530	&    0.843   &	-10.0   $\pm$	0.9	&	151.8	$\pm$	4.0	\\
    2460293.371297	&    0.902   &	-13.8   $\pm$	1.2	&	140.5	$\pm$	4.1	\\
    2460293.379051	&    0.961   &	-16.2   $\pm$	1.1	&	122.2	$\pm$	2.8	\\
	                &            &				        &	                    \\
    2460323.506512	&    0.425   &	-40.4   $\pm$	2.9	&	94.5	$\pm$	1.8	\\
    2460323.514279	&    0.483   &	-41.8   $\pm$	2.1	&	87.2	$\pm$	1.7	\\
    2460323.522045	&    0.542   &	-35.1   $\pm$	1.2	&	91.2	$\pm$	1.7	\\
    2460323.529810	&    0.601   &	-36.1   $\pm$	1.5	&	84.9	$\pm$	2.7	\\
	                &            &				        &			            \\
    2460325.503746	&    0.570   &	-6.7	$\pm$	2.3	&	94.0	$\pm$	2.0	\\
    2460325.511501	&    0.629   &	-2.6	$\pm$	3.4	&	88.4	$\pm$	3.4	\\
    2460325.519267	&    0.688   &	-21.3   $\pm$	5.0	&	68.0	$\pm$	4.7	\\
    2460325.527010	&    0.746   &	0.4	    $\pm$	4.2	&	96.0	$\pm$	11.4\\
	                &            &				        &		                \\
    2460330.489833	&    0.381   &	24.2	$\pm$	3.3	&	121.3	$\pm$	2.1	\\
    2460330.497588	&    0.439   &	-17.0   $\pm$	3.5	&	98.7	$\pm$	1.4	\\
    2460330.505342	&    0.498   &	-60.9   $\pm$	4.0	&	71.9	$\pm$	1.4	\\
    2460330.513096	&    0.557   &	30.9	$\pm$	2.7	&	97.3	$\pm$	2.3	\\
	                &            &				        &			            \\
    2460341.458632	&    0.559   &	-22.8   $\pm$	2.3	&	88.9	$\pm$	3.3	\\
    2460341.466399  &    0.618   &	-27.1   $\pm$	4.0	&	92.4	$\pm$	4.5	\\
    2460341.474175  &    0.677   &	-39.2   $\pm$	2.9	&	94.5	$\pm$	3.3	\\
    2460341.481942	&    0.736   &	-18.5   $\pm$	2.8	&	119.6	$\pm$	4.2	\\
	                &            &				        &		                \\
    2460389.322768	&    0.524   &	-5.5	$\pm$	3.0	&	108.0	$\pm$	4.2	\\
    2460389.330521	&    0.583   &	9.9	    $\pm$	2.4	&	126.9	$\pm$	4.7	\\
    2460389.338288	&    0.642   &	-1.1	$\pm$	2.7	&	123.9	$\pm$	4.5	\\
    2460389.346041	&    0.701   &	6.6	    $\pm$	3.0	&	142.1	$\pm$	5.0	\\
	                &            &				        &		                \\
    2460399.300144  &    0.185   &	-32.5   $\pm$	7.2	&	100.4	$\pm$	4.3	\\
    2460399.307899	&    0.244   &	-57.8   $\pm$	9.0	&	87.3	$\pm$	4.3	\\
    2460399.315676	&    0.303   &	-29.2   $\pm$	15.2&	97.5	$\pm$	6.8	\\
    2460399.323442	&    0.361   &	-23.0   $\pm$	22.1&	93.9	$\pm$	6.4	\\
	                &            &				        &			            \\
    2460403.286812  &    0.417   &	-64.8   $\pm$	16.5&	91.0	$\pm$	5.1	\\
    2460403.294578	&    0.475   &	-86.8   $\pm$	20.0&	80.2	$\pm$	5.9	\\
    2460403.302355	&    0.534	 &  -13.1	$\pm$	24.2&	96.3    $\pm$	18.3\\
    2460403.310109  &    0.593   &   14.1	$\pm$	24.2&	114.7	$\pm$	26.8\\
	                  &            &			          &			              \\
    2460405.283607	&    0.559   &	7.0	    $\pm$	13.9&	96.7	$\pm$	15.6\\
    2460405.291373	&    0.618   &	11.8	$\pm$	5.5	&	123.4	$\pm$	12.5\\
    2460405.299138	&    0.676   &	-3.0    $\pm$	3.3	&	128.5	$\pm$	6.6	\\
    2460405.306904	&    0.735   &	9.1	    $\pm$	2.1	&	178.6	$\pm$	6.2	\\

	\hline
	\hline
\end{tabular}
\end{threeparttable}
\end{table}

\begin{table}[h!]
\centering 
\caption{SALT HRS phase-folded binned radial velocity measurements} 
\label{tab:2SXPS_J0623_HRS_rv_phase_folded}
\begin{threeparttable}
\begin{tabular}{ccc}
\hline
\hline
$\phi$    & H$\alpha$ Blue    & H$\alpha$ Red    \\
(mid bin) & (km s$^{-1}$)  & (km s$^{-1}$)  \\
    \hline
    0.05	&	23.4     $\pm$	0.6	&	170.8   $\pm$	2.2	\\
    0.15	&	-11.9    $\pm$	0.7	&	103.6   $\pm$	2.4	\\
    0.25	&	2.4	     $\pm$	1.9	&	89.7	$\pm$	4.0	\\
    0.35	&	19.3	 $\pm$	3.1	&	105.9   $\pm$	2.3	\\
    0.45	&	-39.8    $\pm$	2.1	&	84.3	$\pm$	1.2	\\
    0.55	&	-20.0    $\pm$	2.5	&	87.0	$\pm$	2.5	\\
    0.65	&	-21.8    $\pm$	2.2	&	87.0	$\pm$	2.7	\\
    0.75	&	-7.0	 $\pm$	1.3	&	126.5   $\pm$	3.6	\\
    0.85	&	-12.9    $\pm$	0.8	&	120.4   $\pm$	3.2	\\
    0.95	&	-9.0	 $\pm$	0.8	&	116.7   $\pm$	2.2	\\
	\hline
	\hline
\end{tabular}
\end{threeparttable}
\end{table}

\clearpage
    
\section{Photometry} \label{App:Photometry}

\begin{figure}[h!]
        \centering
        \includegraphics[width = 1.0\textwidth]{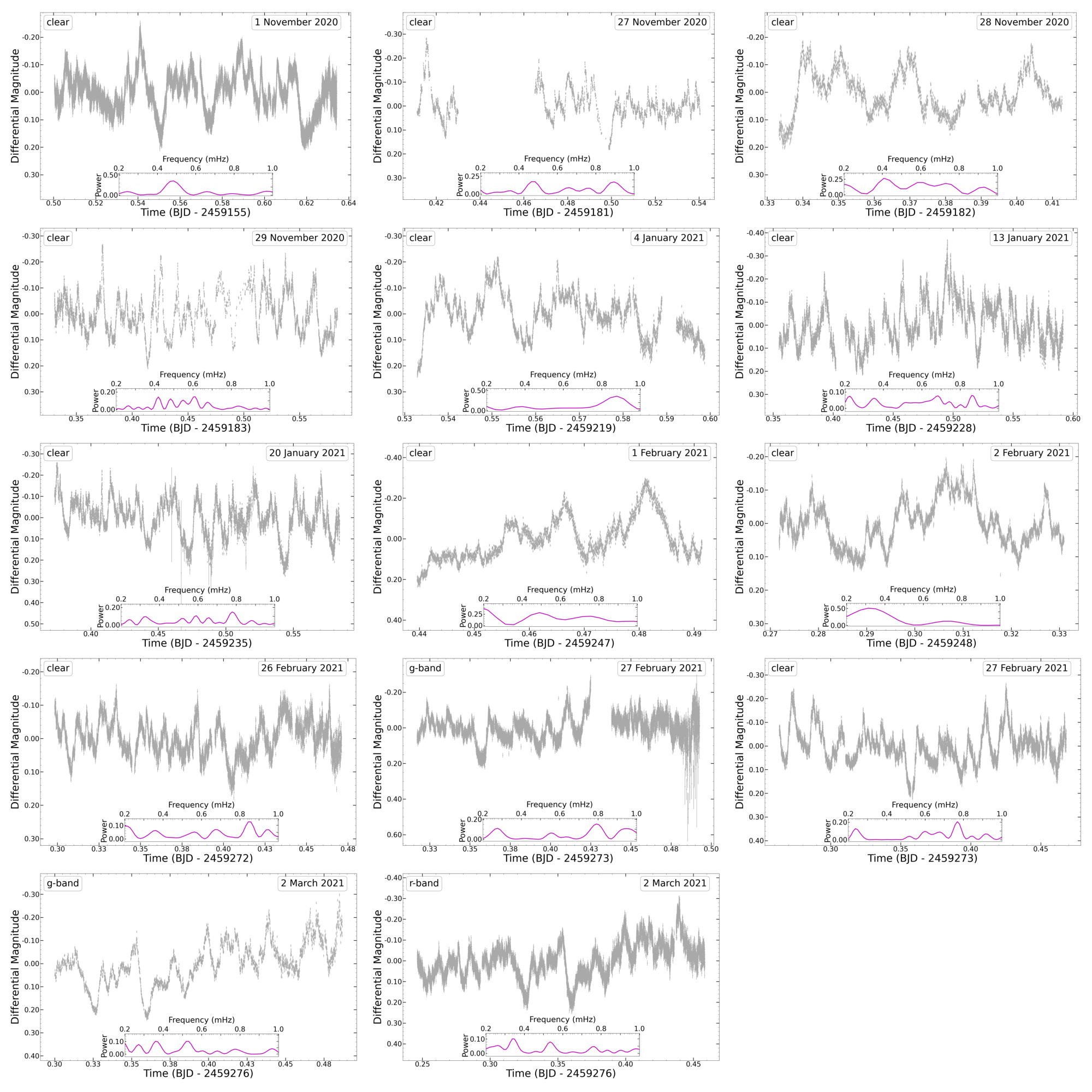}
        \caption[Photometry]{Light curve plots form all photometric observations. The filter used for the observations is found in the top left corner of each respective figure. The inset plot shows the power spectrum from Lomb-Scragle period analysis.}
        \label{fig:2SXPS_J0623_lightcurve_set}
\end{figure}

\clearpage

\begin{table}[htb!]
\centering 
\caption{Differential photometry periods} 
\label{tab:2SXPS_J0623_differential_photometry_all_periods}
\begin{threeparttable}
\begin{tabular}{lccc}
\hline
\hline
Date    & Filter    & Orbits        & Period    \\
        &           & Covered       & (hours)   \\
    \hline
	2020 November 1     & clear & 1.015	    &  0.580  $\pm$ 0.006 \\   
	2020 November 27    & clear & 0.980	    &  0.586  $\pm$ 0.022 \\  
	2020 November 28    & clear & 0.602	    &  0.670  $\pm$ 0.055 \\  
	2020 November 29    & clear & 1.922	    &  0.457  $\pm$ 0.101 \\   
	2021 January 4      & clear & 0.499	    &  0.317  $\pm$ 0.035 \\   
	2021 January 13     & clear & 1.795	    &  4.524  $\pm$ 0.212 \\   
	2021 January 20     & clear & 1.600 	&  0.356  $\pm$ 0.030 \\   
    2021 February 1     & clear & 0.393     &  1.556  $\pm$ 0.039 \\
    2021 February 2     & clear & 0.448     &  0.858  $\pm$ 0.028 \\
    2021 February 26    & clear & 1.310     &  2.896  $\pm$ 0.036 \\
    2021 February 27    & $g'$  & 1.332     &  3.378  $\pm$ 0.026 \\
    2021 February 27    & clear & 1.554     &  0.361  $\pm$ 0.086 \\
    2021 March 2        & $g'$  & 1.417     &  5.606  $\pm$ 0.069 \\
    2021 March 2        & $r'$  & 1.609     &  6.365  $\pm$ 0.328 \\
	\hline
	\hline
\end{tabular}
\end{threeparttable}
\end{table}

\begin{figure}[h]
\includegraphics[width =\columnwidth]{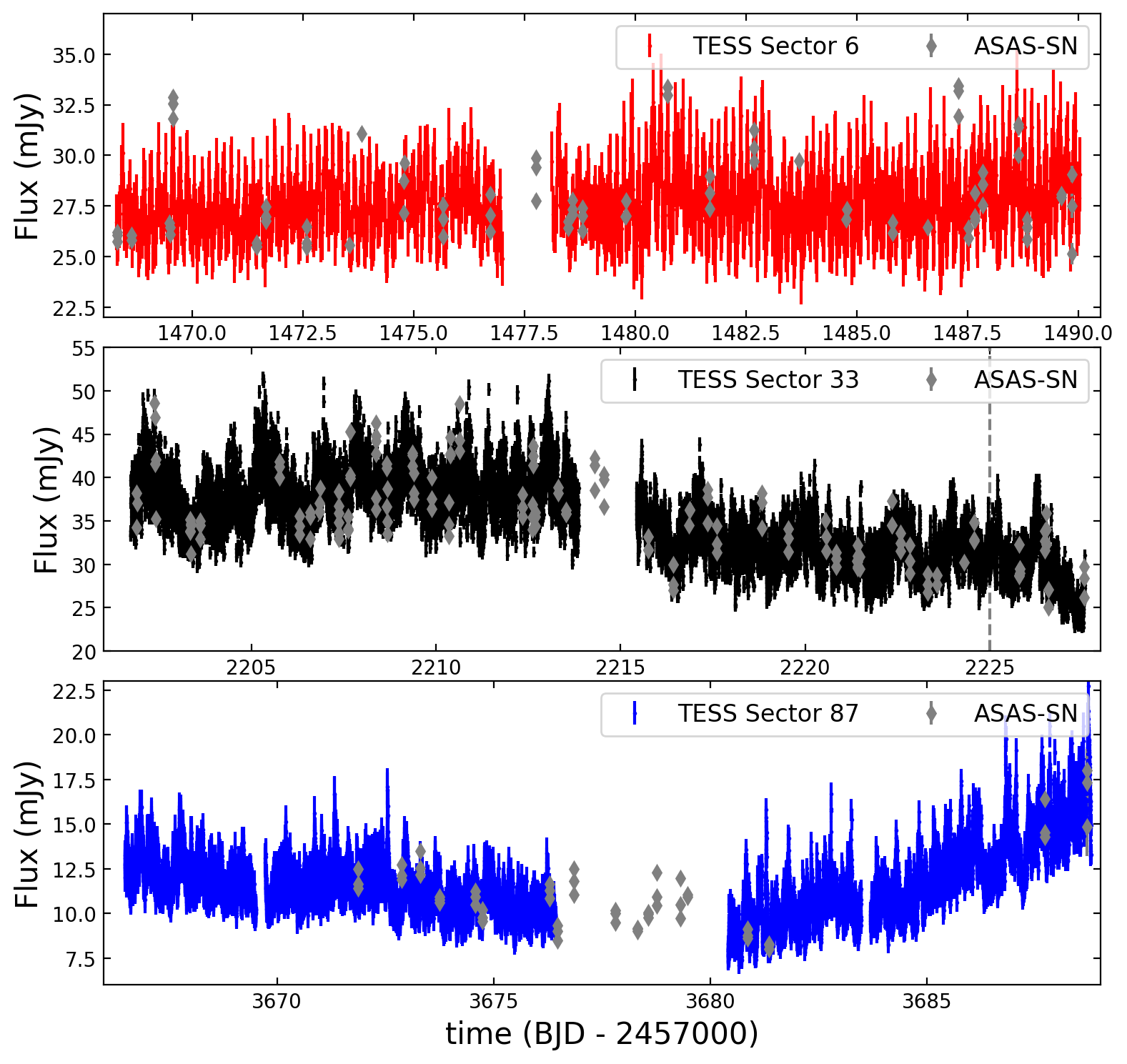}
\centering
\caption[\textit{TESS} Sector 6, 33 and 87 light curves calibrated to ]{\textit{TESS} Sector 6 (top), Sector 33 (middle), and Sector 87 (bottom) light curves. The grey diamonds represent \asassn\ $g$-band observations. The vertical dashed grey line in the middle panel shows a cut-off point, after which the light curve was no longer stationary, and hence excluded from analysis.}
\label{fig:2SXPS_J0623_TESS_lightcurves}
\end{figure} 

\begin{figure}[h]
        \centering
        \includegraphics[width =\columnwidth]{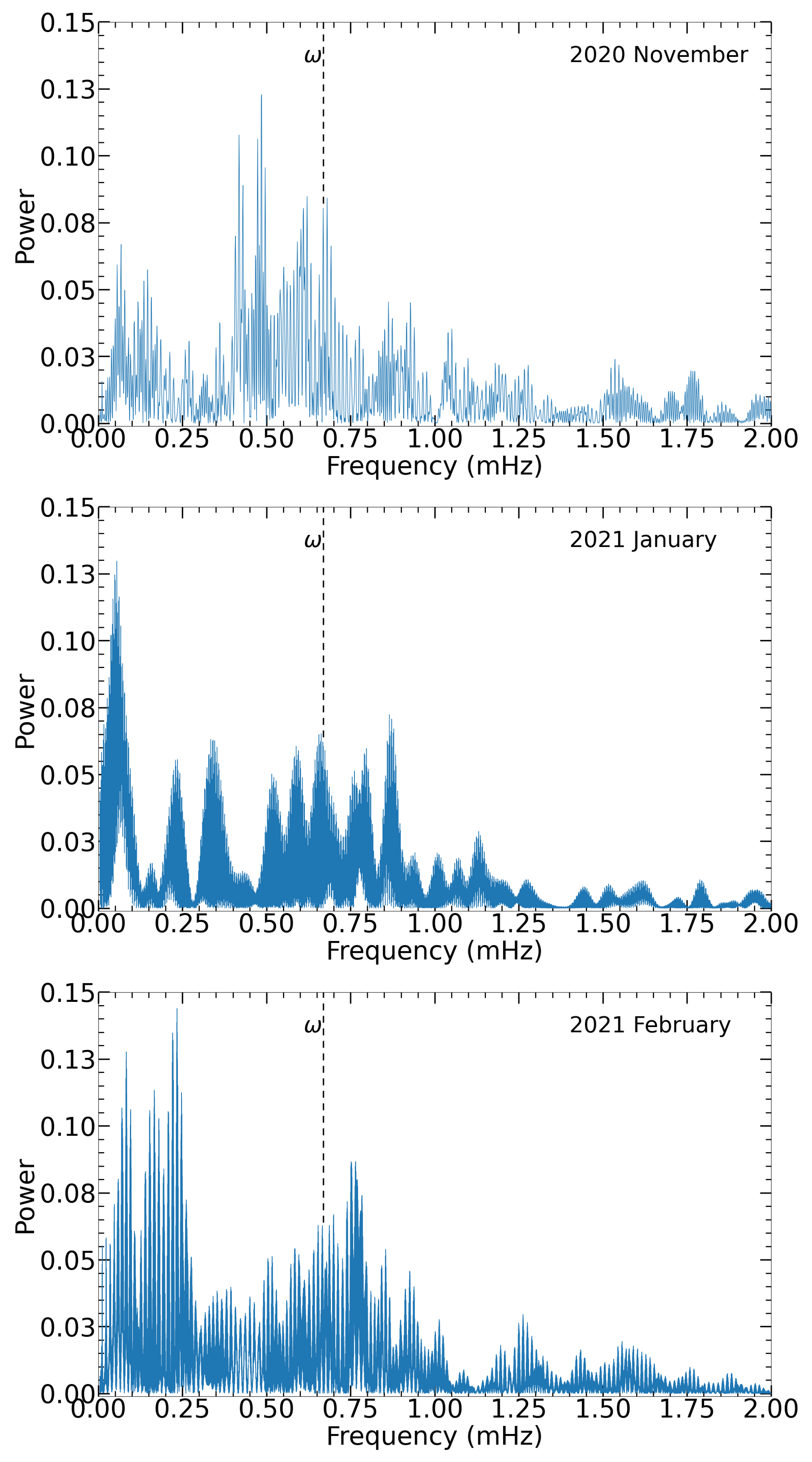}
        \caption[grouped photometry power spectrum]{Lomb-Scargle power spectra of the clear filter photometric observations. The observations were divided into three groups, the combined observations obtained in 2020 November (top panel), those obtained in 2021 January (middle panel) and those obtained in 2021 February (bottom panel). In each panel, the spin frequency of the WD is indicated by the dashed line labelled $\omega$, and show that some power is seen in each grouping at $\omega$. It is evident from these three groupings that there is significant change in the power spectrum on short (weeks, or shorter) timescales, which could be interpreted as showing evidence of QPOs.}
        \label{fig:2SXPS_J0623_month_photometry_power_spectrum}
\end{figure}

\begin{figure}[h]
        \centering
        \includegraphics[width=\columnwidth]{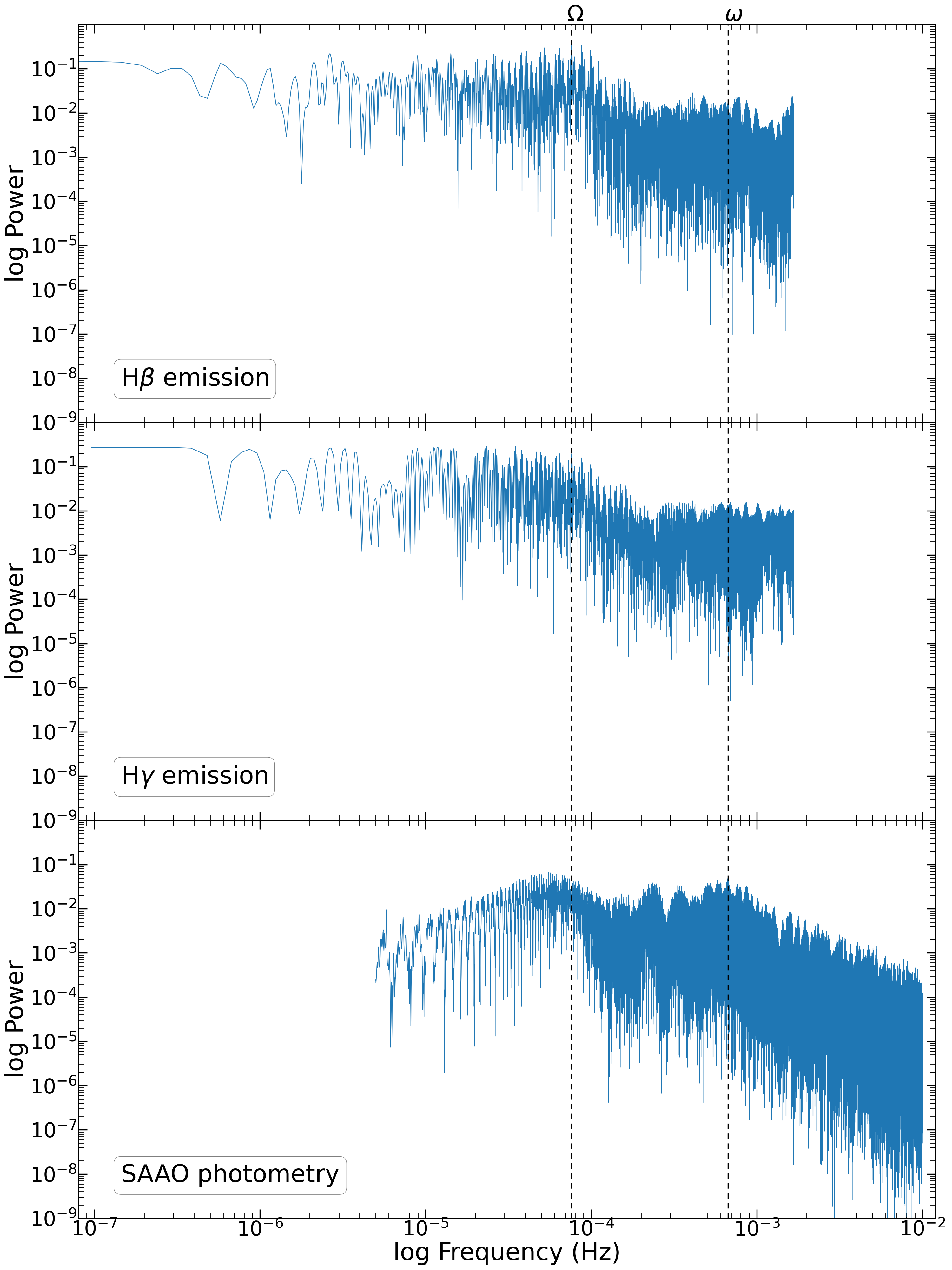}
        \caption[log-log power spectra]{Log-log power spectra of the H$\beta$ emission line radial velocities (top), H$a\gamma$  emission (middle), and the SAAO photometric observations (bottom). Each panel extends to the respective Nyquist frequency of the observation. The dashed vertical lines indicate the spectroscopic orbital, $\Omega$, and spin, $\omega$, frequencies, respectively.}
        \label{fig:SGRt_0623_rv_power_density_spectra_all_data}
\end{figure}

\end{appendix}

\end{document}